\documentclass[apj,twocolumn]{emulateapj}
\usepackage{natbib,lscape}
\shorttitle{XPSs and Radio Galaxies in Clusters of Galaxies}
\shortauthors{Hart, Stocke \& Hallman}

\def\radiolimit{$3\times10^{23}$ W Hz$^{-1}$ }
\def\xraylimit{$10^{42}$ ergs s$^{-1}$ }
\def\xrayunits{ergs~s$^{-1}$}
\def\radiounits{W Hz$^{-1}$}

\begin{document}
\title{X-ray Point Sources and Radio Galaxies in Clusters of Galaxies}
\author{Quyen N. Hart, John T. Stocke, and Eric J. Hallman}
\affil{Center for Astrophysics and Space Astronomy, \\Department of Astrophysical and Planetary Sciences, 
\\UCB-389, University of Colorado, Boulder, CO 80309}

% ********************** ABSTRACT *******************************
\begin{abstract}
Using {\it Chandra} imaging spectroscopy and VLA L-band maps, we have identified radio galaxies at
P$_{(1.4 GHz)}\geq$ \radiolimit and X-ray point sources (XPSs) at L$_{(0.3-8 keV)} \geq$ \xraylimit 
in 11 moderate redshift ($0.2<z<0.4$) clusters of galaxies.  Each cluster is uniquely 
chosen to have a total mass similar to predicted progenitors of the present-day Coma Cluster.
Within a projected radius of 1 Mpc we detect 20 radio galaxies and 8 XPSs (3 sources are detected in both X-ray and radio) 
confirmed to be cluster members above these limits.  75\% of these are detected within 500 kpc (projected) of the cluster center.
This result is inconsistent with a random selection from bright, red sequence ellipticals at the $>$ 99.999\% level.  We
suggest that these AGN are triggered somehow by the intra-cluster medium (ICM), perhaps similar to the Bondi accretion model of 
\citet{2006MNRAS.372...21A}.  All but one of the XPSs are hosted by luminous ellipticals which otherwise show no other evidence 
for AGN activity.  These objects are unlikely to be highly obscured AGN since there is no evidence for large amounts of X-ray or optical absorption.
One XPS, in addition to possessing a pure absorption-line optical spectrum, has a large excess of light blueward 
of the Ca II H\&K break that could be non-thermal emission; a second XPS host galaxy probably has excess blue light.
The most viable model for these sources are low luminosity BL Lac Objects, similar to the high-energy-peaked BL Lacs (HBLs)
discovered in abundance in serendipitous X-ray surveys.
The expected numbers of lower luminosity FR 1 radio galaxies and HBLs in our sample converge to suggest that very 
deep radio and X-ray images of rich clusters will detect AGN (either X-ray or radio emitting or both) in a large fraction of 
bright elliptical galaxies in the inner 500 kpc.
Because both the radio galaxies and the XPSs possess relativistic jets, they (and, by extension, the entire RLF) can inject 
heat into the ICM. Using the most recent scalings of P$_{jet} \propto$ L$_r^{0.5}$ from \citet{2008ApJ...686..859B}, radio
sources weaker than our luminosity limit probably contribute the majority of the heat to the ICM.  Also, because
these heat sources move around the cluster, AGN heating is distributed rather evenly.
If a majority of ICM heating is due to large numbers of low power radio sources, triggered into activity by the increasing ICM density
as they move inward, this may be the feedback mechanism necessary to stabilize cooling in cluster cores. 

\end{abstract}

% ********************************  Introduction **************************************
\section{Introduction}
\label{sec:intro}
Observations of galaxy clusters have provided important clues about cosmology, structure
formation and the evolution of galaxies.  As
the largest gravitationally bound objects in the Universe, clusters are unique
locations to study the creation and evolution of AGN. Based upon the observed correlation between 
supermassive Black Hole mass and galaxy bulge mass 
\citep{1998AJ....115.2285M}, we expect that there should be numerous, luminous AGN in the 
bright ellipticals in rich clusters which can affect their surroundings. 
Indeed, observational evidence supports the notion that non-gravitational processes have affected the entropy of the 
intracluster medium (ICM), as suggested by the steepness of the L$_x$-T$_x$ relationship compared to self-similar 
scaling expectations \citep{1991MNRAS.252..414E, 1998ApJ...504...27M, 2004ApJ...611..158R}. Radio-loud
AGN have been directly implicated in cluster ICM heating because evacuated ``bubbles'' in the diffuse X-ray emitting gas 
have been observed that are spatially coincident with non-thermal radio emission 
\citep{2002MNRAS.331..369F, 2004ApJ...607..800B}. However, up until now only AGN in the brightest cluster galaxies 
(BCGs) have been observed to spawn these bubbles and so only the central AGN has been incorporated into ICM heating models 
\citep[e.g.,][]{2005ApJ...630..740B}. In this case, theoretical difficulties arise when attempting to distribute 
this heating throughout the central cluster regions \citep[e.g.,][]{2007ApJ...671..171V}.
A natural solution would be to have other cluster AGN injecting energy throughout the
inner cluster region \citep[e.g.,][]{2006MNRAS.373..739N}.

Most cluster AGN studies have either employed clusters in a flux-limited survey 
\citep[e.g.,][]{1999AJ....117.1967S,2005ApJ...623L..81R,2006AN....327..571B} or simply selected clusters based upon availability
\citep[e.g.,][]{2007ApJ...664..761M}. But since flux-limited surveys produce cluster samples
which contain more luminous and thus more massive objects at higher redshifts, this selection naturally
identifies more evolved structures at earlier times.  This is clearly opposite to the type of selection 
one would prefer to use to study cluster, galaxy or AGN evolution. Moreover, this mismatch can result in 
comparisons that can obscure any true evolution, since more massive clusters are placed at the 
beginning of the evolutionary sequence and less massive objects at the end.

In this paper, we choose galaxy clusters by using a new selection method that avoids the
difficulty just described.  In this summary of first results, we
use this sample to investigate the nature of radio galaxies and X-ray point sources (XPSs) in clusters at moderate-$z$ 
(0.2--0.4). For now, 
we avoid using the term X-ray AGN until the AGN nature of the XPSs is proven (or disproven) by our observations presented below.
Previous studies simply assumed that these X-ray emitters are AGN (e.g., the ``optically-dull X-ray AGN''; 
\citealp{2002ApJ...576L.109M,2006ApJ...644..116M,2007ApJ...664L...9E}). 
The AGN nature of these XPSs has not been carefully scrutinized before and their X-ray and radio properties could have a 
significant impact on models for AGN heating in clusters.
We address each of these points in detail in this paper. We describe our sample 
selection in \S~\ref{sec:sample} and data analysis of X-ray and radio selected AGN in \S~\ref{sec:data}. In 
\S~\ref{sec:results} we discuss the observational properties of the radio and X-ray cluster source populations, including location
within the cluster and X-ray-to-radio flux ratios. In \S~\ref{sec:bl_lac} we discuss the nature of the cluster XPSs.  In \S~\ref{sec:discussion}
we discuss the implications of this study, particularly for the cluster heating problem. In \S~\ref{sec:conclusions} we summarize our results and
suggest testable predictions.  We use H$_0$ = 70 km/sec, $\Omega_{\Lambda}=0.70$, and $\Omega_{M}=0.3$ throughout. 
While we present a summary of our general methodologies and detected sources here, a more detailed
discussion will be presented at a later time in \citet{hart_future}.

% ********************************  Road to Coma Description **************************
\section{The ``Road to Coma'' Sample}
\label{sec:sample}

\begin{figure*}
\begin{center}
\includegraphics{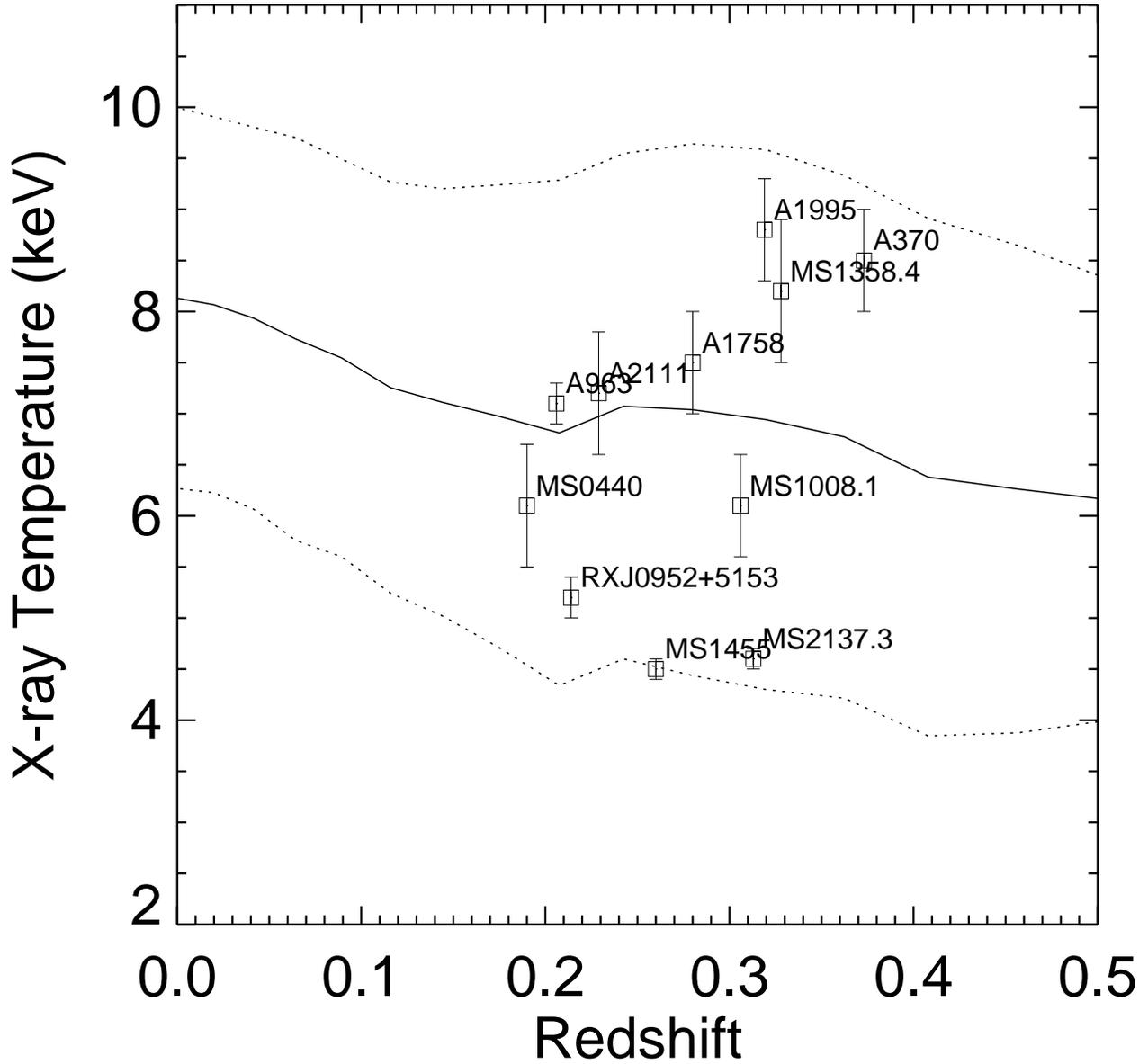}
\caption[scale=0.5]{\scriptsize{
Predicted X-ray temperature (keV) versus redshift for Coma Cluster progenitors.
The model curves show results from detailed hydrodynamical simulations
(similar to \citealt{1988MNRAS.235..911E} and \citealt{2001ApJ...555..597B}) of the 
evolution of clusters which evolve to masses similar to the Coma Cluster.  
Measures of $T_X$(z)/T$_X$(z=0) from the simulations
are normalized by the Coma Cluster's present temperature to illustrate the variance of cluster
properties. The solid line shows the median value for this distribution and the dashed
line shows the 25th and 75th percentiles.  Eleven clusters are chosen to be a
representative sample of Coma Cluster progenitors at moderate redshifts (0.2$<$z$<$0.4). 
See \S~\ref{subsubsec:xray_data_icm} for details on the bulk ICM temperature determination.
}}
\label{fig:road_to_coma}
\end{center}
\end{figure*}

Guided by hydrodynamical simulations to track the growth of massive cluster 
halos ($M>10^{15} M_\sun$) from $z\sim1$ to the present epoch, we have used the 
temperature of the ICM as a proxy for cluster mass.  We  
selected $0.2<z<0.4$ clusters that are predicted to evolve into objects like the 
present-day Coma Cluster (kT=8.2$\pm$0.2; \citealt{2008ApJ...687..968L}).
The details of the cosmological simulations (similar to \citealt{1988MNRAS.235..911E})
which include preheating of the ICM \citep{2001ApJ...555..597B} will be fully described 
in \citet{hart_future}.  In short, measurements of $T_X$(z)/T$_X$(z=0) for
individual massive cluster halos in the simulation are normalized by the 
Coma Cluster's present temperature to illustrate the variance of cluster properties.
Figure~\ref{fig:road_to_coma} displays the median temperature of these simulated Coma Cluster progenitors 
as a function of redshift (solid line) and the 25th and 75th percentiles about this distribution (dashed lines).
At z$=$0.3, the ICM temperature of a Coma Cluster progenitor is predicted to be 7.0$\pm$2.6 keV.  The
temperature spread is due partly to the modest number of realizations of Coma Cluster mass-scale objects in these simulations.
At z$=$1.0, the permitted temperature range for our sample slowly decreases to 4.0$\pm$2.0 keV. 
Our unique method attempts to select clusters of similar mass at a given redshift and 
thus these clusters on the ``Road to Coma'' will be used for consistent evolutionary comparisons later.

We selected clusters readily available in the {\it Chandra} archives with adequate exposure time to
detect X-ray sources with $L_{0.3-8.0 keV}\sim1\times10^{42}$ ergs~s$^{-1}$.  This luminosity is well above
the expected emission from low-mass X-ray binaries in ellipticals \citep[L$_{X}<5\times10^{38}$ \xrayunits;][]{2004ApJ...611..846K}
and high-mass X-ray binaries in starburst galaxies \citep[L$_{X}<5\times10^{40}$ \xrayunits;][]{2003MNRAS.339..793G}.  Also, for this redshift range, a 1 Mpc
radius region falls on one ACIS chip and so has been selected consistently as our survey radius
centered on the peak of the diffuse X-ray emission.  
We also require radio observations to detect radio galaxies with P$_{1.4GHz} \geq$~3~$\times$
10$^{23}$ W~Hz$^{-1}$ across the entire survey.  Along with our well-defined cluster
sample, our multi-wavelength approach to identify cluster AGN  
differs from many previous studies \citep[e.g.,][]{1995AJ....109..853L,2007ApJ...664..761M} which use 
X-ray or radio observations, but not both, to detect these objects
in only one waveband. Table~\ref{tab:data_obs} lists our eleven $z$=0.2-0.4 
``Road to Coma'' clusters, their redshifts, {\it Chandra} observational details (ObsID, ACIS aimpoint, 
exposure time, ICM temperature, flux and luminosity limits), as well as VLA 1.4 GHz radio details.  
The following section details the X-ray, radio and optical observations and data analysis.

% ********************************  Data Analysis **************************
\section{Multi-wavelength Data Analysis}
\label{sec:data}
% ********************************  X-ray Data Analysis **************************
\subsection{X-ray Imaging Spectroscopy}
\label{subsec:xray_data}
% *************************************************
We re-processed the {\it Chandra} archival observations using CIAO v3.3 and CALDB v2.2.
Following the typical ACIS data preparation pipeline, event-1 files were reprocessed to 
remove bad pixels, afterglow pixels, and streaking patterns.  The newest calibration
files for charge transfer inefficiency and time-dependent gain were applied, and then finally 
the event files were filtered on status and grade.  Flaring events were identified by extracting the 0.3-12 keV count rate
on either the ACIS-S1 or ACIS-I1 chips and we eliminated time periods with rates
$>$3$\sigma$ above the observed mean.

\subsubsection{Cluster ICM Temperature}
\label{subsubsec:xray_data_icm}
Cluster ICM spectra were extracted from within a 1 Mpc projected radius after obvious point sources
were removed (point source identification is described below).  If a background area 
could not be extracted on the same chip as the cluster emission, we extracted
background spectra from re-projected blank sky observations. Obvious cool-core regions were
excised prior to spectral modeling.  Source spectra were binned to a minimum of 20 cts/bin, then modeled
using {\it XSPEC} \citep{1996ASPC..101...17A} and MEKAL models with foreground extinction held fixed at 
the Galactic hydrogen column at the cluster coordinates \citep{1990ARA&A..28..215D} and a constant 0.3 Solar metal abundance. 

The lower redshift limit of our cluster sample is limited by the size of one ACIS chip, allowing
a $\sim$~1 Mpc radius to fall on one chip.  A 1 Mpc radius corresponds to $\sim$~0.5 R$_{200}$.
However, our simulated cluster temperatures were extracted from within R$_{200}$.
Based on the similarity in the temperature profiles of many clusters for r$>$0.1 R$_{200}$, in both cool-core and non-cool cores
\citep[e.g.,][]{2002ApJ...567..163D,2005ApJ...628..655V,2007ApJ...666..835B,2007A&A...461...71P}, 
we expect the ICM temperature profile to be decreasing beyond 0.5 R$_{200}$. Thus, our temperature estimates 
extracted within a 1 Mpc projected radius will slightly overestimate the full cluster temperature 
by $\sim$~10--20\% relative to those derived in the simulations to select Coma-like clusters \citep{2001ApJ...555..597B}.
Therefore, the T$_x$ values listed in column 6 of Table~\ref{tab:data_obs} need to be decreased by 
10--20\% to be accurately compared to the simulated Coma Cluster progenitor temperatures mentioned in the previous section.

\subsubsection{X-ray Point Sources (XPSs)}
\label{subsubsec:xray_data_xps}
XPSs were identified using the CIAO tool {\it wavdetect} 
with a threshold limit set to $10^{-6}$ which corresponds to one false source detection per $10^6$ pixels
(roughly the number of pixels in an unbinned image of one ACIS chip). Each potential source above this threshold was scrutinized 
individually. Broadband (0.3-8.0 keV) counts were extracted within a 95\% encircled energy radius estimated 
for a monochromatic energy source of 1.5 keV at the observed off-axis angle.  Background annuli regions were 
2.0 times the extraction radius.  This background subtraction does not account for the slow change of
T$_x$ with radius within a cluster.  However, a more sophisticated background subtraction using annuli at the
XPS cluster radius show resulting differences less than the quoted photon statistics

Conversion from net counts to flux was estimated for the XPSs using XSPEC assuming a 
power-law spectrum with photon index of $\Gamma$=1.7 (N$_{E}\propto E^{-\Gamma}$). 
To estimate the X-ray flux limit for each observation (Column 7 of Table~\ref{tab:data_obs}), 
we estimated the point source flux for a 3$\sigma$ detection above the background noise near
the survey edge (R $=$ 1 Mpc).  Abell 2111 has a formal 3$\sigma$ detection limit 
of $L_{0.3-8.0 keV}\sim1.3\times10^{42}$ ergs~s$^{-1}$, slightly greater than our stated survey 
threshold of $L_{0.3-8.0 keV}\geq10^{42}$ ergs~s$^{-1}$.  However, we
feel comfortable leaving this cluster in our sample because the CIAO {\it wavdetect} routine identified a point 
source with $L_{0.3-8.0 keV}\sim8\times10^{41}$ ergs~s$^{-1}$ if the source is located at the cluster redshift.
Poissonian errors were calculated using the \citet{1986ApJ...303..336G}
approximation.  X-ray sources with SNR$>$3 and with expected K-corrected, rest-frame $L_{0.3-8.0 keV}\geq1\times10^{42}$ ergs~s$^{-1}$
are catalogued as potential cluster AGN candidates.  
We note that standard detection methods described above may miss XPSs associated with the BCG due to misidentifying AGN
emission as unresolved ``cool core'' ICM emission; i.e., the cusp of the ICM emission can mask the presence of an AGN.
Therefore, we can make no solid claims about X-ray AGN in the bulk of the individual BCGs (see \S~\ref{sec:results} 
for a further discussion.)

% ********************************  Radio Data Analysis **************************
\subsection{Radio Imaging}
\label{subsec:radio_data}
% *************************************************
Several of our low-$z$ clusters fields have been surveyed by the Faint Images of the Radio Sky at 20cm \citep[FIRST; ][]{1995ApJ...450..559B} 
and/or the NRAO VLA Sky Survey \citep[NVSS; ][]{1998AJ....115.1693C} to detect radio galaxies with $P_{1.4GHz}\geq3\times10^{23}$ Wm$^{-2}$. 
For our redshift range, the resolution of the FIRST maps may underestimate the total radio power of galaxies if any extended
flux is close to the noise limit of the maps.  
Conversely, two radio sources with small angular separation may be catalogued as one object by FIRST.  
% To mitigate these issues, we obtained additional 20-cm VLA maps (VLA program ID AS873, mostly in C-array), 
To mitigate these issues, we obtained additional 1.4 GHz VLA maps (Program ID AS873) and
re-analyzed 1.4~GHz VLA archival observations obtained in its A-configuration.

% Specifics of AS873 Observations
In September/December 2006 we obtained VLA continuum observations at 1.4 GHz (50 MHz bandwidth) for four clusters 
(MS 0440.5$+$0204, Abell 2111, MS 1455.0$+$2232, and Abell 1995).  Abell 1995 was observed for 75 minutes in B-array, 
while the remaining three low-z clusters were observed for 30 minutes in C-array.
The duration of a typical target scan was 15 minutes, bracketed by 1 minute scans of a nearby
phase calibrator.  A flux calibrator (e.g. 3C286) was observed for 3 minutes at the beginning and end of each observing session.

We used the NRAO Astronomical Imaging Processing System (AIPS), version 31DEC07, in the usual manner 
to flag, calibrate, transform and clean the images, so that radio sources could be detected manually. 
Source flux densities are estimated with the AIPS task TVWIN and IMEAN.
The typical 3$\sigma$ limit of our observations and FIRST is 0.3--0.5 mJy and 0.4 mJy, respectively.
For Abell 963, the 3$\sigma$ limit of the FIRST image is higher ($\sim$~1.1 mJy) due to a bright background radio source with
S$_{1.4GHz}\sim$~1.4~Jy, but the observations still result in a radio power limit below our required threshold.  
Higher resolution A-array maps for three clusters (MS 0440.5$+$0204, MS 2137.3-2353, Abell 370)
were reduced in a similar manner.  In some instances, these higher resolution maps reveal two distinct radio sources for a 
single FIRST detection.
%(e.g., MS 0440.5+0204 R3 \& R4, Abell 370 R1 \& R3).  
While these higher resolution maps were used to provide the best optical 
identifications and to resolve close pairs of sources, the lower resolution maps 
(FIRST or other B/C-configuration maps) were used for flux determinations. 

In Table~\ref{tab:data_obs} columns 10-11 list the 1.4 GHz flux density and K-corrected luminosity limits for the individual clusters at the
edge of our 1 Mpc survey region (K-correction assumes F$_{\nu}\propto\nu^{-\alpha}$ and $\alpha$=0.7).
These limits allow detection of many lower radio power FR 1 sources while excluding the lower luminosity radio sources due to
star formation \citep[L$_{1.4GHz}<10^{22.75}$W Hz$^{-1}$;][]{2003ApJS..146..267M}.  
Given the very low radio power limits in column 11, Table~\ref{tab:data_obs}, 
we do not expect that we have missed sources due to the presence of strong point sources in or near the primary beam, 
partially resolving the source or to beam-smearing in the outer regions of the fields, excepting MS1358.4+6245.
Due to a strong, nearby radio source in this field, \citet{1999AJ....117.1967S} only detected the BCG at 1.4 GHz.
This 1.4 GHz map has a larger radio power limit (P$_{1.4GHz}\geq9.7\times10^{23}$ \radiounits) for potential cluster radio galaxies 
than the other clusters in our sample.  However, in a 5 GHz map \citet{1999AJ....117.1967S} did not detect any additional 
sources within 5' of the BCG down to a 3$\sigma$ limit of 0.2 mJy, which corresponds to 1.4 GHz flux limit of 0.5 mJy and
P$_{1.4GHz}\geq1.7\times10^{23}$ \radiounits.  Thus, we include MS1358.4+6245 in our cluster sample with only one detected radio source in the BCG.

% ********************************  Optical Data Analysis **************************
\subsection{Optical Datasets}
\label{subsec:opt_data}
% *************************************************

% ********************************  Optical Data Analysis **************************
\subsubsection{Multi-band Imaging}
\label{subsec:opt_imaging}
% *************************************************
Two-color images and photometry are publicly available from the the Sloan Digital Sky Survey Data Release 6 
\citep[SDSS DR6;][]{2008ApJS..175..297A}, the Canadian Network for Observational Cosmology images \citep[CNOC;][]{1996ApJS..102..269Y}, and/or 
the {\it Chandra} Multi-wavelength Project \citep[ChaMP;][]{2004ApJS..150...43G} for 10 of our 11
clusters in this paper.  

For 6 of our 11 clusters, SDSS DR6 provides five color imaging (\emph{ugriz}) and photometry. 
The SDSS limiting magnitude in (\emph{g,r,i}) is (22.2,22.2,21.3), adequately deep to detect M$>$M$_{r}^{*}$
cluster galaxies at z=0.4.  SDSS photometric uncertainty is 1\% \citep{2008ApJS..175..297A}.
The CNOC cluster redshift survey examined 16 X-ray luminous clusters with 0.2$<$z$<$0.55.
For 3 of our 11 clusters, we utilized CNOC two-color photometric (Gunn g,r) and spectroscopic catalogs that are based on
MOS observations from the Canada-France-Hawaii Telescope (CFHT) between 1993--1994 (see \citealt{1996ApJS..102..269Y} for
details).  For z$<$0.45 the CNOC limiting Gunn-r magnitude reaches down to M$_{r}^{*}+$1 with an individual source magnitude uncertainty
of $\sim$~0.3 mag.  ChaMP is a wide-area survey of serendipitous X-ray sources in $>$100 archival {\it Chandra} observations.
For MS 2137.3-2353, we use ChaMP's multi-band optical images (Sloan \emph{g,r,i}), obtained at NOAO 4 m with the Mosaic CCD cameras
and publicly available at http://hea-www.cfa.harvard.edu/CHAMP/CHAMP.html.  
% The limiting r-magnitude of the ChaMP images of MS 2137.3-2353 is 25.2 (see \citealt{2004ApJS..150...43G}).

In November 2007 we obtained Sloan (\emph{g,r,i}) images of Abell 370 with SPIcam on the Astrophysical Research Consortium (ARC) 3.5m Telescope
at Apache Point Observatory (APO).  The FOV of SPIcam is 4.8 arcmin$^2$; therefore, we required multiple
pointings of Abell 370 to image the entire 1 Mpc radius region (FOV $\sim$~6.5 arcmin$^2$).  We used NOAO's Image Reduction and Analysis Facility
(IRAF), v2.14.1, in the usual manner to bias subtract and flat field individual images.  Images were aligned and stacked
into a large mosaic image with the IRAF package MSCRED.  The limiting r-band magnitude ($\sim$~22.4) reaches down to 
M$_{r}^{*}$+1 with an individual source magnitude uncertainty of $\sim$~0.3 mag.
For APO and ChaMP images, we used the Picture Processing Program \citep[PPP; ][]{1991PASP..103..396Y} to detect and classify 
objects into stars and galaxies.  PPP is a robust detection algorithm, especially in crowded fields, such as
galaxy clusters.  Additionally, total source magnitudes are calculated using optimal extraction radii that are based
on photometric growth curves.  

% ********************************  Optical Data Analysis **************************
\subsubsection{Spectroscopy}
\label{subsec:opt_spec}
% *************************************************

SDSS and/or CNOC spectroscopic data are available for several candidate cluster XPSs and radio galaxies.
The SDSS spectral coverage is 3800-9200 \AA\ with a resolution of 1800-2200.
The SDSS spectroscopic pipeline (spectro1D) cross-correlates individual wavelength and flux-calibrated spectra with
various template spectra (stellar, quasar, and emission-line galaxies) to determine source redshifts 
via emission lines and/or absorption features (see \citealt{2002AJ....123..485S} for details). 
The CNOC spectral coverage is 4650-6100 \AA\ with a dispersion of $\sim$~3.45\AA/pixel, yielding a 
spectral resolution of 16.5 \AA\ with a 1.5\arcsec\ slit width.  The CNOC survey also incorporates
cross-correlation techniques to determine source redshifts that have velocity uncertainties
between 100--130 km s$^{-1}$ (see \citealt{1996ApJS..102..269Y} for details).  In addition to these two surveys, we
searched the literature (via NASA's Astrophysics Data System Bibliographic Services) and extragalactic databases 
(e.g. NASA/IPAC Extragalactic Database) for published redshifts and/or spectra of our potential cluster targets.

For remaining cluster AGN candidates without published redshift and/or spectra, spectroscopic observations were obtained 
for all objects with Sloan $r<20.8$ using the Double Imaging Spectrograph (DIS-II) at the APO 3.5m between 2006--2009. 
Observations were obtained in its low resolution ($\sim$~6 \AA) mode which provides complete spectral coverage from 3700--9000 \AA.
Comparison spectra (He, Ne, Ar) were routinely obtained between targets and/or long exposure sets and subsequently were used to calibrate
the wavelength scale of science spectra.  Exposure times ranged from 45 minutes to 2.5 hours, achieving a SNR~$>$ 5 per 
pixel in the science spectra.  
Spectroscopic images were reduced with IRAF in the usual manner (bias subtraction and flat fielding) and source
spectra were extracted and calibrated with the IRAF APEXTRACT package.

Source redshifts were calculated from individual wavelength measurements of typical absorption features 
(e.g. the 4000\AA~Ca break, G-band, Mg Ib, Na I) and/or emission-line features (e.g. [O~II], [O~III], H$\alpha$).
The radial velocity accuracy from a typical spectral line is $\sim$~300 km s$^{-1}$ (or $\Delta$z~$\sim$~0.001).  
At this writing, spectra for radio galaxies and XPSs are complete to $M_r<-20.8$ 
($M^{*}_r\sim-20.8$ for typical SDSS DR4 rich cluster galaxies; \citealt{2005ApJ...633..122H}).
We define passive galaxy spectra as those showing only 
stellar absorption lines (e.g., the 4000\AA~Ca break, Mg Ib, Na I), while active galaxy spectra have detectable emission lines 
(e.g., [O~II], H$\alpha$, [O~III]).  We do not find any example of a starburst optical 
spectrum among either the radio galaxies or XPSs as determined by comparisons of strong emission-line luminosities
(e.g. [O~III] 5007 \AA\ comparable to ~H$\beta$; see \citealt{1991ApJS...76..813S}).  This method is indicative, and 
not as robust as ``BPT diagnostic plots" \citep{1981PASP...93....5B} that utilize several emission-line features to 
separate starburst galaxies from AGN.

% *************************************************
\subsubsection{Cluster Red Sequence Galaxies}
\label{subsubsec:crs_calc}
% *************************************************
The color-magnitude diagrams (CMDs) of clusters generally display a tight color sequence for the
cluster ellipticals, referred to as the cluster red sequence (CRS).  We estimate the number of CRS galaxies 
within 1 Mpc of the cluster center and
with M$_r\leq-20.8$ (where our spectroscopy becomes incomplete) in the following manner.
Foreground and background galaxies are statistically subtracted from each cluster, using
Sloan r-band background counts from \citet{2001AJ....122.1104Y}.
We determine the mean CRS color by fitting a bimodal Gaussian distribution
to both the red and blue galaxy populations as a function of color.  Using only galaxies
with colors within $3\sigma$ of the mean CRS color, we define the CRS more precisely by fitting a line to the observed color and magnitude
of these sources and thus verifying the well-known ''tilt" of the CRS for cluster color-magnitude diagrams
\citep[e.g.,][]{1977ApJ...216..214V}.
For galaxies with M$_r\leq-20.8$, the large majority of blue galaxy contaminators are foreground galaxies and not the
fainter blue cloud galaxies with M$_r>M^{*}_{r}$.  
The estimated number of CRS galaxies will be used to compute AGN fractions described in \S~\ref{subsec:radial_distribution}.

% ********************************  RESULTS ***********************************************
\section{Results}
\label{sec:results}
% ******************************************************************************************

Within 1 Mpc of the cluster X-ray emission centroid,
we find 8 XPSs with $L_{0.3-8.0 keV}>1\times10^{42}$ ergs~s$^{-1}$, confirmed to be 
cluster members within the $3\sigma$ velocity dispersion of the cluster (from the literature) and $M_{r}<-20.8$,
with the exception of Abell 963 X2 with M$_r=-20.3$, which we include despite being 0.5 mag fainter. 
We also find 20 radio galaxies with $P_{1.4GHz}>3 \times 10^{23}$ W Hz$^{-1}$ and $M_{r}<-20.8$,
of which six are located in the BCG. 

Tables~\ref{tab:xps}--\ref{tab:radio_galaxies} lists the basic data for the XPSs and radio galaxies, respectively, in our
sample.  The sources are identified by cluster name and a consecutive numbering which includes foreground and background sources.
In Table~\ref{tab:xps} the letter identifiers in parentheses are used in subsequent plots and discussions for easy cross-referencing. 
Both tables include the spectroscopic redshift (mostly
from this work), observed magnitude of the host galaxies in which they are found, the projected radial 
distance from the cluster center as defined by the diffuse ICM X-ray emission peak in h$^{-1}_{70}$ Mpc (called ''radius"),
and observed broadband (0.3-8 keV) X-ray luminosity or limit.  Table~\ref{tab:xps} also includes the net broadband (0.3--8.0 keV)
counts for each XPS.

Table~\ref{tab:radio_galaxies} also includes the radio source flux density at 1.4 GHz and K-corrected 1.4 GHz radio luminosity.
Host galaxy colors and cluster ``red sequence'' (CRS) colors (see \S~\ref{subsubsec:crs_calc}) 
were obtained using Sloan (\emph{g-r}) for z$<$0.3 clusters or (\emph{r-i}) 
for $z$=0.3--0.4 clusters (photometric calibration errors are typically 2\% in these colors).  For the 3 EMSS clusters, CNOC photometry
provides Gunn (g-r) colors.  We note that MS 0440.5+0204 R3 \& R4 and Abell 370 R1 \& R3 are examples of two distinct sources
detected in the higher resolution VLA maps, but were identified as a single source in FIRST images.

For radio galaxies with detectable X-ray emission, the X-ray luminosities (Table~\ref{tab:radio_galaxies}, column 12)
were determined in the same manner as for the detected XPSs described in \S~\ref{subsubsec:xray_data_xps}.
The upper X-ray luminosity limits for the BCGs and radio galaxies with X-ray non-detections above our limits
are estimates as 3$\sigma$ above the ICM emission at their locations.
For Abell 1758, the ICM X-ray emission is both diffuse, allowing a good estimate for the upper limit on
the X-ray luminosity of the BCG, and double-peaked, indicative of a merging system.
In this case we have determined the projected distance of the radio galaxies to be from the nearest diffuse ICM
emission center. Thus, A1758-R4 is identified with the BCG in the NW X-ray clump and A1758-R6 is 2/3 Mpc from the other clump center
to the SE. Otherwise, there are no ambiguous cases in the projected distance (in Mpc) from the center of the ICM X-ray emission.
The radio upper limits for the XPS locations are set at 3$\sigma$ above the background noise and so are conservative.

Due to the difficulty of detecting XPS at the center of the diffuse X-ray emission, we probably have 
underestimated the number of cluster XPSs.  We did detect one BCG in both radio and X-rays, Abell 1758 R4/X1 -- listed
in Tables~\ref{tab:xps}--\ref{tab:radio_galaxies}, where the ICM emission is particularly diffuse.
Since 6 of 11 BCGs were detected as radio galaxies (refer to Table~\ref{tab:radio_galaxies}),
we might expect that a similar fraction of BCGs are XPSs.  However, since our radio and X-ray detection limits
allow the detection of only half the total number of XPSs as radio galaxies (refer to Table~\ref{tab:xps}) we conservatively
estimate that $\sim$~3 BCGs are XPSs at $L_{0.3-8.0 keV}>$~10$^{42}$ ergs~s$^{-1}$ in our sample, one of which we have already identified (A1758 R4/X1).

\begin{figure*}
\begin{center}
\includegraphics{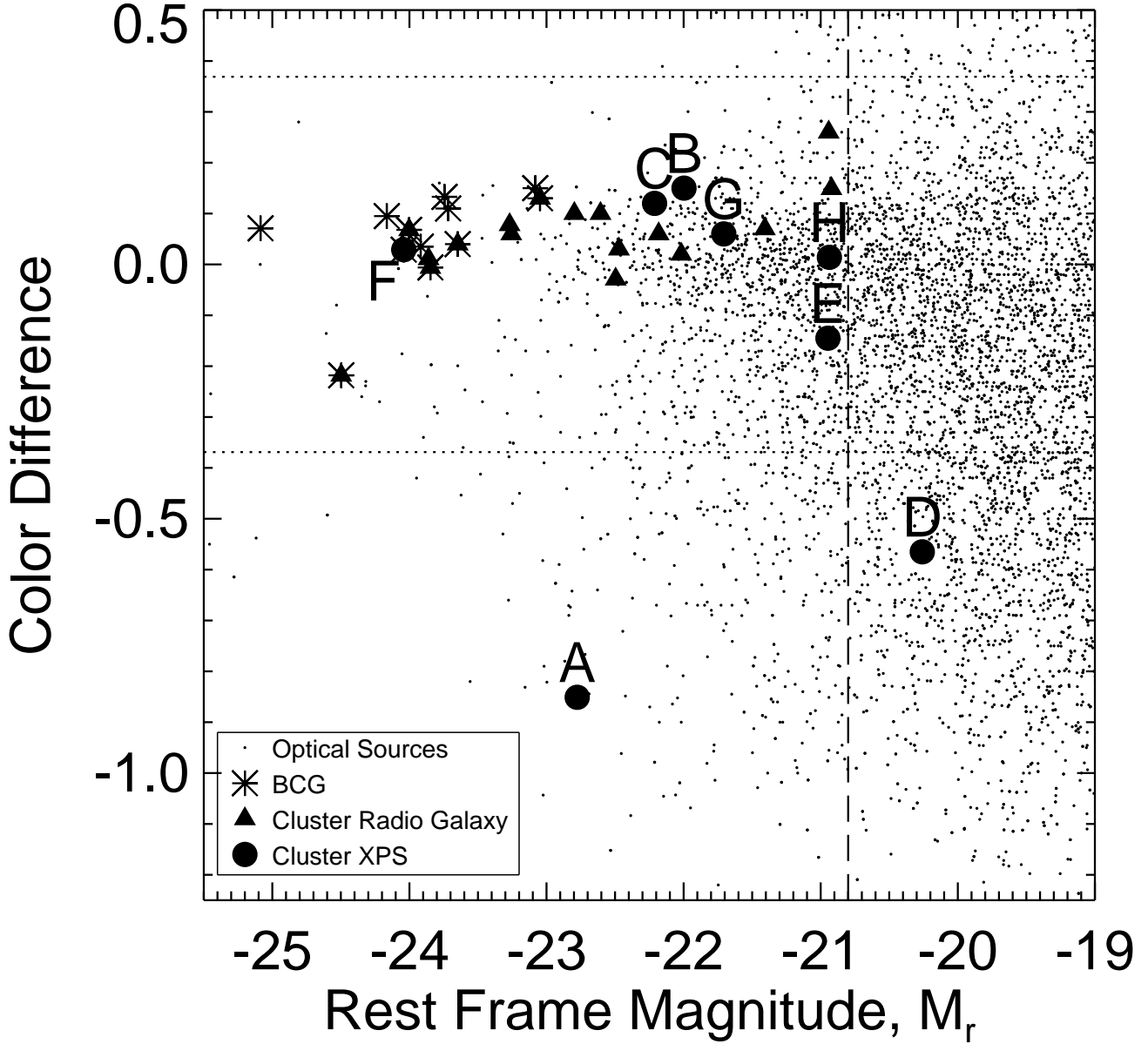}
\caption[scale=0.5]{\scriptsize{
Composite color-magnitude diagram for eleven cluster fields.  For each cluster, the color difference is
computed relative to the average color of the cluster red sequence galaxies (see \S~\ref{subsubsec:crs_calc})
All optical sources within the 1 Mpc projected radius are plotted with small dots.
Brightest cluster galaxies (BCGs) are identified by an asterisk (with two BCGs for Abell 1758 which appears double-peaked in
X-rays).  Notice that cluster radio galaxies (filled triangles) and cluster
 X-ray point sources (filled circles) are located along the cluster
red sequence, implying that they are hosted by passive, elliptical galaxies.
Letter identifiers corresponds to X-ray point sources listed in Table~\ref{tab:xps}.
M$^*$=~-20.8 is marked by a vertical dashed line.  The horizontal dotted lines
are representative of the 3$\sigma$ spread ($\sim$~0.36 mag) from the mean CRS color over the entire cluster sample.
In general the cluster red sequence has a slope of $\sim$~-0.03 based on SDSS (g,r)
photometry of clusters in our sample, with more luminous CRS galaxies having redder colors.
}}
\label{fig:cmd}
\end{center}
\end{figure*}

Figure~\ref{fig:cmd} displays a composite color-magnitude diagram for our eleven low-$z$ clusters.  
All but one radio galaxy lies on the CRS.
The exception is the BCG in MS 1455.0+2232 for which extremely luminous, extended [O~II] 3727 \AA\ emission is present in the 
g-band, making it appear to be somewhat bluer than the CRS \citep{1992ApJ...385...49D}. Also excepting one case (Abell 963 X1, object A in
Figure~\ref{fig:cmd}), XPSs are hosted by early-type galaxies on or near the CRS. Abell 963 X1 has an optical spectrum consistent 
with a Seyfert nucleus (with emission-line luminosity of [OIII]$>>$H$\beta$).  Given the presence of only one 
Seyfert projected within 1 Mpc of a cluster core and $\sim$~745 km s$^{-1}$ relative radial velocity for this Seyfert
(Abell 963 cluster velocity dispersion of $\sigma_v$=1350 km s$^{-1}$; \citealt{1999ApJS..125...35S}), 
this source is consistent with being a cluster member either foreground or background to the core 
at a physical radius $>$ 1 Mpc.  A second XPS (Abell 963 X2, object D in Figure~\ref{fig:cmd}), 
the least luminous XPS host galaxy in our sample, is several tenths of a 
magnitude bluer than the red sequence, but, nonetheless, possesses a passive absorption line spectrum. This galaxy will be discussed in
detail in \S~\ref{subsec:lum_ratio}.

% *************************************************
\subsection{AGN Fraction and Radial Distribution of Cluster Red Sequence (CRS) Radio Galaxies and X-ray Point Sources} 
\label{subsec:radial_distribution}
% *************************************************

Table~\ref{tab:agn_frac} lists our estimates for the number of background-subtracted CRS galaxies
for each cluster (see \S~\ref{subsubsec:crs_calc} for details), the total number of cluster radio and X-ray sources,
the radio and X-ray source fraction, color used to determine the CRS number, and the optical survey utilized.
Background-subtracted CRS galaxies number $\sim$~665 in our entire cluster sample.
The average AGN fraction for CRS galaxies in our sample is 3\% for radio-selected sources and 1\% for X-ray selected sources within 1 Mpc
of the cluster center.

\begin{figure}
\begin{center}
\includegraphics{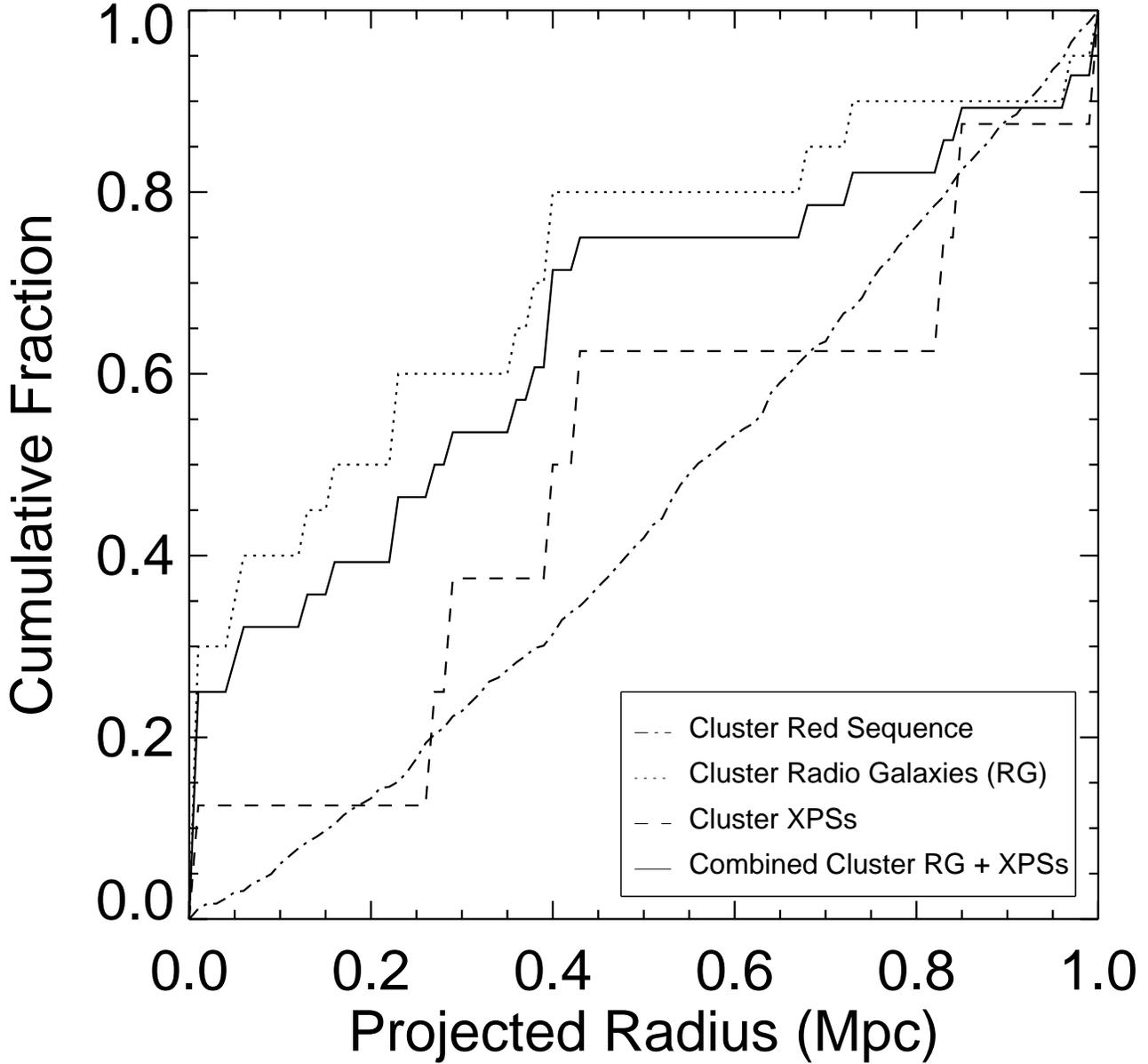}
\caption[scale=0.5]{\scriptsize{
The cumulative fraction of cluster red sequence (CRS) galaxies ({\it dot-dash line}), cluster radio galaxies ({\it dotted line}),
cluster X-ray point sources (XPSs) ({\it dashed line}), and combined cluster radio galaxies and XPSs ({\it solid line}).
The population of cluster radio galaxies and XPSs are more centrally concentrated than cluster red sequence galaxies with
$\sim$~75\% of them located within a 500 kpc projected radius from the cluster X-ray centroid.  A Kolmogorov-Smirnov
test rules out at the 99.999$\%$ level that the radio galaxies and XPSs are drawn from the CRS parent population, suggesting
that AGN triggering is connected to the ICM environment.
}}
\label{fig:radial_dist}
\end{center}
\end{figure}

The cumulative radial distributions of the cluster radio galaxies and XPSs, as shown in Figure~\ref{fig:radial_dist},
reveal these sources to be more centrally-concentrated in the cluster relative to CRS galaxies as a whole.
Eighty percent of radio galaxies and 60\% of XPSs are located within 500 kpc of the cluster centers
and appear more centrally concentrated than the CRS galaxies even in that region.  
A luminosity weighting of the CRS distribution using the bivariate radio luminosity function derived by \citet{1977A&A....57...41A}
is indistinguishable from the unweighted CRS galaxy distribution; i.e., there is no evidence for significant mass segregation in these clusters. 
Thus, the significant difference between the radio galaxy $+$ XPS population projected radial distances and the weighted CRS galaxy 
radial distances remains as shown in Figure~\ref{fig:radial_dist}.

% Added a couple of sentences on additional two XPS at r=0
A two-sided Kolmogorov-Smirnov (K-S) test between the full X-ray $+$ radio population and the CRS galaxy population as a whole 
are inconsistent with being drawn from the same parent population at $>$ 99.999\% confidence level (K-S D-statistic = 0.36 and 
probability = 3.3$\times$10$^{-6}$).  The K-S probability is even lower if we only examine the cluster radio sources
(K-S D-statistic = 0.44 and probability = 4.5$\times$10$^{-9}$).  The K-S probability is noticeably larger for just the cluster
XPSs (K-S D-statistic = 0.26 and probability = 1.9$\times$10$^{-3}$), but this comparison is still significant at 
the 99.8\% confidence level.  To account for the likely presence of XPSs in BCGs which are not detected due to 
the presence of an ICM ``cool core'' cusp in the diffuse X-ray emission, we add two XPSs at r=0.  The K-S probability
for this augmented XPS distribution (K-S D-statistic = 0.38 and probability = 7.1$\times$10$^{-7}$) is 
comparable to the original X-ray $+$ radio population and only strengthens our conclusion of a centrally concentrated
population of XPSs.  

Within 500 kpc 16 radio galaxies in our sample make up 6$\pm$2\% of the bright (L$\gtrsim L^*)\sim$~265 CRS galaxies (40\% of 665), 
while the 4 XPSs (excluding Abell 963 X1) make up 2$\pm$1\%.  The combined radio and XPS population make up 
8$\pm$2\% of the bright (L$\gtrsim L^*)\sim$~265 CRS galaxies.
Our result strongly suggests that the triggering of a radio-loud
AGN or an XPS in these cluster galaxies is due to some, as yet undetermined, mechanism related to the ICM density such
as in the Bondi accretion model of \citet{2006MNRAS.372...21A}.

% *************************************************
\subsection{X-ray Properties}
\label{subsec:xray_properties}
% *************************************************
Of the 8 detected XPSs (Table~\ref{tab:xps}), we detect 7 XPSs with $25-115$ net broadband (0.3-8.0 keV) counts 
and one XPS (Abell 963 X1) with $\sim$~2150 net counts. This XPS is the bright (M$_r\sim$-22.8), blue (almost 1
magnitude bluer than the CRS) galaxy in Figure~\ref{fig:cmd} (Object A) and possesses a typical Seyfert optical spectrum. 
Using {\it XSPEC}, the spectrum of Abell 963 X1 is well-modeled by a
power-law with spectral index $\Gamma=2.0\pm0.1$, emission from the 6.4 keV 
Fe K-$\alpha$ line with an equivalent width of 1.7 keV, and no intrinsic absorption. But this spectrum also can be fit by a 
Raymond-Smith thermal model with kT = 3.4$^{+0.5}_{-0.3}$ keV, but at a higher reduced-$\chi^2$ value (1.18 vs. 1.03). 
Given the clear Seyfert-like optical spectrum and likely power-law X-ray spectrum we identify Abell 963 X1 as a clear AGN.

But due to low photon counts in the other 7 XPSs, we instead compute X-ray colors to estimate their 
spectral slope and amount of intrinsic absorption. 
Figure~\ref{fig:xray_color} displays the ratio of two X-ray colors. The three bands (S1, S2, and H) are defined for
similar energy ranges as in \citet{2004ApJ...600...59K} and \citet{2006ApJ...644..116M}.
Overlaid are the expected colors for a power-law spectrum with spectral slopes of $\Gamma=1.0-4.0$ and intrinsic 
$n_H=0,10^{21},10^{22},10^{23}$ cm$^{-2}$. Each individual source is displayed and labeled as in Figure~\ref{fig:cmd} and in Table~\ref{tab:xps}; 
the composite X-ray color (filled square) is also displayed using all the XPSs without detectable radio emission and the XPSs with detectable
radio emission (filled triangle), excepting Abell 963 X1 (Object A). 
With the exception of MS0440 X8 (Object G) and RXJ0952 X6 (Object H), the XPSs are consistent with a typical 
AGN photon index of $\Gamma\sim1.75$ \citep{2006A&A...451..457T}. MS0440 X8 has relatively few hard 
(2.5-8.0 keV) counts and has consistent colors for source with $\Gamma \sim 3$, but as with the others, little evidence 
for obscuration in the X-rays.  RXJ0952 X6 has approximately equal counts in the soft (0.5-2.5 keV) and
hard (2.5-8.0 keV) bands and has consistent colors for a source with $\Gamma \sim 1$.  

In addition to the analysis above, a composite X-ray spectrum in five discrete rest-frame bands is 
consistent with a power-law model with $\Gamma=1.6\pm0.2$ with no evidence for intrinsic absorption.
Thus, if these XPSs are AGN, they are {\it unobscured} AGN. There is no evidence for X-ray spectral differences for XPSs 
with and without radio detections, and, although the statistics are modest, this evidence also weighs in favor of all of these
XPSs being AGN of some sort.

% ********************************  X-ray/Radio Luminosity Limits ********************************
\subsection{X-ray to Radio Luminosity Ratios}
\label{subsec:lum_ratio}
% *************************************************
We find that XPSs and radio galaxies rarely overlap in our sample. 
There are only 3 XPSs which are radio-detected, suggesting two populations. 
Are we really seeing two populations of sources or is this just a function of our flux limits? To investigate 
this further, we determined the X-ray to Radio Luminosity ratios for our cluster sources (Figure~\ref{fig:lum_ratio})
and compared them to optical elliptical ``core'' galaxies in \citet{2006A&A...447...97B}, more luminous 3C/FR 1 radio galaxies 
in \citet{2004ApJ...617..915D} and Seyferts hosted by early-type galaxies in \citet{2007A&A...469...75C}.
A loose linear relationship exists in Figure~\ref{fig:lum_ratio} for the 
low (optical ``core'' galaxies) and higher power (FR 1s) ellipticals of the
\citet{2006A&A...447...97B} and \citet{2004ApJ...617..915D} samples, respectively. The X-ray limits on the 
radio galaxies in our sample are consistent with other FR 1s; i.e., we would require perhaps a factor of $\sim$~3 
improvement in X-ray sensitivity before we would expect to detect all of these radio galaxies.

On the other hand, the XPSs in our sample would be easily detected in the radio {\it if} they were 
similar to FR 1 radio galaxies; indeed, they should have been detected at power levels comparable to or significantly higher
than the radio galaxy detections we did make. So, we can rule out that the radio galaxies and the XPSs are drawn from the same
parent population. But the XPSs do
have X-ray luminosities and radio luminosity limits that place them among the Seyferts in Figure~\ref{fig:lum_ratio}. 
If these XPSs are some sort
of Seyfert galaxy, an order of magnitude improvement in our radio flux limit probably would be required 
to detect all of them. Indeed, if the XPSs are all Seyferts, it is somewhat surprising that we did not detect some of them already
given the distribution of Seyferts in Figure~\ref{fig:lum_ratio}.

Are the XPSs simply radio-quiet AGN, similar to Seyfert galaxies? While they possess
X-ray luminosities of Seyferts, they do not possess the 
emission-line spectra of Seyferts nor is there unambiguous evidence for obscuration. 
Even if we use the X-ray to H$\alpha$ flux ratios (mean value = 7.3) of much lower luminosity AGN 
like in the nearby sample of \citet{2008ARA&A..46..475H}, we would expect H$\alpha$ and other emission line fluxes for our sample of XPSs at 
L$_{H\alpha}>$ 10$^{41}$ ergs s$^{-1}$, which we easily exclude using our 3.5m spectra. We have also gone through the exercise of
determining if the Seyfert spectrum (i.e., blue power-law and luminous emission lines of $[O III]$ and H$\alpha$) of Abell 963 X1 would 
be detectable if scaled down by the difference in X-ray luminosity between this sources and the others (using standard scaling relations 
observed for X-ray emitting Seyferts).
We find that a Seyfert-like AGN of power 10\% of Abell 963 X1 would be detected easily in our optical spectra and a 3\% power-level probably
would be detected as well. But it might be possible to ``hide'' a Seyfert 2 type spectrum (line luminosities can be an order of magnitude 
lower than many Seyfert 1s), which leads to the suggestion that these could be
highly-obscured AGN. However, as we showed above, there is no strong evidence for obscuration in these XPSs based on their
X-ray spectra. Nor are their optical broad-band colors significantly redder than the CRS although a low luminosity AGN with 
M$_r \sim -18$ could be hidden in these luminous galaxies without affecting the broad-band colors within our photometric
errors. We conclude that the XPSs are unlikely to be similar to Seyferts or that they are highly-obscured AGN of any description.

% ********************************  COMPARISON ***********************************************
\subsection{Comparison to Other Cluster AGN Studies}
\label{subsec:comparison}
% ******************************************************************************************
How do our results compare to other cluster AGN studies?  Cluster radio galaxies have been known to reside in passive galaxies
\citep{2003ApJS..146..267M}, thus our findings on their host galaxies are not surprising.
Statistical studies of cluster radio galaxies have also revealed enhanced concentration towards cluster
centers \citep[e.g.,][]{1995AJ....109..853L,2007ApJS..170...71L} similar to what we have found.

\citet{1995AJ....109..853L} surveyed $\sim$300 low-z (z$<$0.09) Abell clusters
to find radio galaxies within $\sim$0.6  Mpc of the cluster optical center.  The
authors find that the surface density distribution of radio galaxies with log(P$_{1.4GHZ}$)$>$23.23
is more centrally concentrated than an expected King model distribution of cluster galaxies.  
\citet{1995AJ....109..853L} argue that the central concentration of radio galaxies in clusters is
a direct result of (1) clusters having luminous host galaxies residing in their cores and (2) more luminous 
galaxies having a higher probability of hosting a radio source \citep{1996AJ....112....9L}.  
Our radio galaxy AGN fraction within 0.5 Mpc of the cluster center is $\sim$6\%, consistent with their estimated fraction
($\sim$8\%) of elliptical galaxies hosting radio sources (see Figure 6 in \citealt{1995AJ....109..853L}).

\citet{2007ApJS..170...71L} surveyed 573 X-ray detected clusters that were also observed with NVSS.  The
authors also find the surface density distribution of radio galaxies (P$_{1.4GHZ}$)$>10^{23}$ \radiounits) within 5r$_{200}$
to be more centrally concentrated than Two Micron All-Sky Survey \citep[2MASS;][]{2000AJ....119.2498J}
K-band galaxies with M$_{K}$$<$-24.  They estimate the radio-active fraction (RAF) of cluster and 
field galaxies with P$_{1.4GHZ}>10^{23}$ \radiounits\ and M$_{K}$$<$-24 to be 4.9$\pm$0.5\% and 1.3$\pm$0.5\%, respectively.  
Note for M$_{r}^{*}\sim-20.8$ \citep{2005ApJ...633..122H} and (r-K)$\sim$2.9 for SDSS early-type galaxies \citep{2006MNRAS.366..717C},
then M$_{K}^{*}\sim-23.7$.  These authors surveyed a larger region around clusters (R$\sim$ r$_{200}$) that could account for
the smaller cluster galaxy RAF compared to our value.  Also, there is both a luminosity and a Hubble type bias 
to this difference since many field galaxies are late-type systems which do not harbor radio galaxies.

Studies of X-ray point sources in the fields of small samples of galaxy clusters
\citep[e.g.,][]{2001ApJ...548..624C,2007A&A...462..449B} revealed a slight excess $\sim2\sigma$ in the expected surface density
compared to cluster-free regions. \citet{2005ApJ...623L..81R} surveyed 51 clusters with 0.3$<$z$<$0.7 and
L$_{0.5-2 keV}>$10$^{43}$ \xrayunits\ and found a large excess in the composite surface density of
XPSs within 3.5 Mpc of the cluster center compared to their 20 control fields.
In relaxed clusters, i.e. clusters with smooth, symmetric X-ray emission profiles, \citet{2005ApJ...623L..81R}
find a prominent excess within 0.5 Mpc, whereas clusters with disturbed X-ray morphologies appear to have their XPS excess
distributed more uniformly within the larger 3.5 Mpc survey region.
\citet{2009MNRAS.392.1509G} analyzed 148 galaxy clusters with 0.1$<$z$<$0.9. Within a 1 Mpc
projected region they find a 3$\sigma$ excess in XPS counts compared to 44 control fields, resulting
in $\sim$~1.5 XPSs per cluster across their sample.
\citet{2009MNRAS.392.1509G} find more AGN in the outer 0.5-1.0 Mpc regions than within 0.5 Mpc.
In these larger statistical surveys all XPSs (after a statistical background subtraction) are assumed to be at
the cluster redshift, but little or no redshift information exists to verify cluster membership.

However, \citet{2006ApJ...644..116M, 2007ApJ...664..761M} has obtained redshifts for cluster AGN candidates.
They examined eight low-z clusters (0.05$<$z$<$0.31) with cluster
ICM temperatures ranging from 4 to 16 keV (including MS1008
which overlaps with our sample) and with 3 fields surveyed out to a 1 Mpc projected radius.
Although their X-ray luminosity limit is one order of magnitude lower and their
magnitude completeness limit is one magnitude fainter than M$^{*}_{R}$, their R-band images show that most of their X-ray
point sources are hosted by galaxies with colors (see \citealt{2007ApJ...664..761M}, Fig. 3)
consistent with (B-R)=1.57$\pm$0.2 of local ellipticals \citep{1995PASP..107..945F}.
So while \citet{2006ApJ...644..116M, 2007ApJ...664..761M} do find a few blue cluster AGN with
Seyfert-like spectra (2 out of 6 X-ray sources with L$_X>10^{42}$ and located within 1 Mpc of the cluster cores),
their numbers are small.
If we assume that the XPSs in our sample are all AGN, then our X-ray AGN fraction of $\sim$2\% is consistent with 
\citet{2006ApJ...644..116M} for $L_{0.3-8.0 keV}>$~10$^{42}$ \xrayunits.
It is interesting to note that although we have a more robust selection method
for our low-$z$ cluster sample, the radio galaxy AGN fraction of \citet{1995AJ....109..853L}
and XPS AGN fraction of \citet{2007ApJ...664..761M} are similar to our values.

% *************************************************
\section{The Nature of Cluster X-ray Point Sources (XPSs)}
\label{sec:bl_lac}
% *************************************************

As stated in the previous section, our cluster XPSs do not appear to have optical nor X-ray signatures typical of Seyferts.
However, the XPSs have many of the properties of low luminosity BL Lac Objects of the class now called ``High-energy-Peak'' 
BL Lacs (HBL; \citealt{1995ApJ...444..567P}; first called X-ray selected BL Lac Objects; \citealt{1985ApJ...298..619S}). 
With X-ray luminosities $\sim$~10$^{42}$ ergs s$^{-1}$ an HBL would be expected to have a radio luminosity of 10$^{21-23}$ 
W Hz$^{-1}$ based upon HBLs at higher total luminosities \citep[e.g.,][]{1996ApJ...456..451P},
at or below our radio detection limits for this sample. Indeed, 3 of these XPSs are radio-detected. 
Most HBLs previously studied have X-ray and 
radio luminosities 2--3 orders of magnitude more luminous than our detections and still {\it only sometimes} exhibit the featureless 
power-law optical spectrum of classical BL Lacs \citep{1999ApJ...516..145R}. Mostly, HBLs have optical spectra of passive elliptical galaxies with
emission line luminosities $<$ 10$^{40}$ ergs s$^{-1}$, consistent with what we find for the cluster XPSs. 

\citet{1999ApJ...516..145R} investigated the general properties of low-luminosity HBLs, finding that their 
properties merge with normal ellipticals, with normal (i.e., $\sim$~50\%) Ca II H \& K ``break strengths'' seen 
in non-AGN ellipticals. \citet*{1995PASP..107..803U} reviewed the other properties of these objects 
which find them in giant (M$_r$= -22 to -25) elliptical galaxies \citep{1997ApJ...480..547W,1999Ap&SS.269..647F}; 
and in rich groups to moderately rich clusters at similar
redshifts to our sample \citep{1997ApJ...480..547W}. HBLs show no evidence for internal X-ray obscuration 
and some show quite soft X-ray spectra like MS0440 X8 \citep{1996ApJ...456..451P}.

\begin{figure}
\begin{center}
\includegraphics{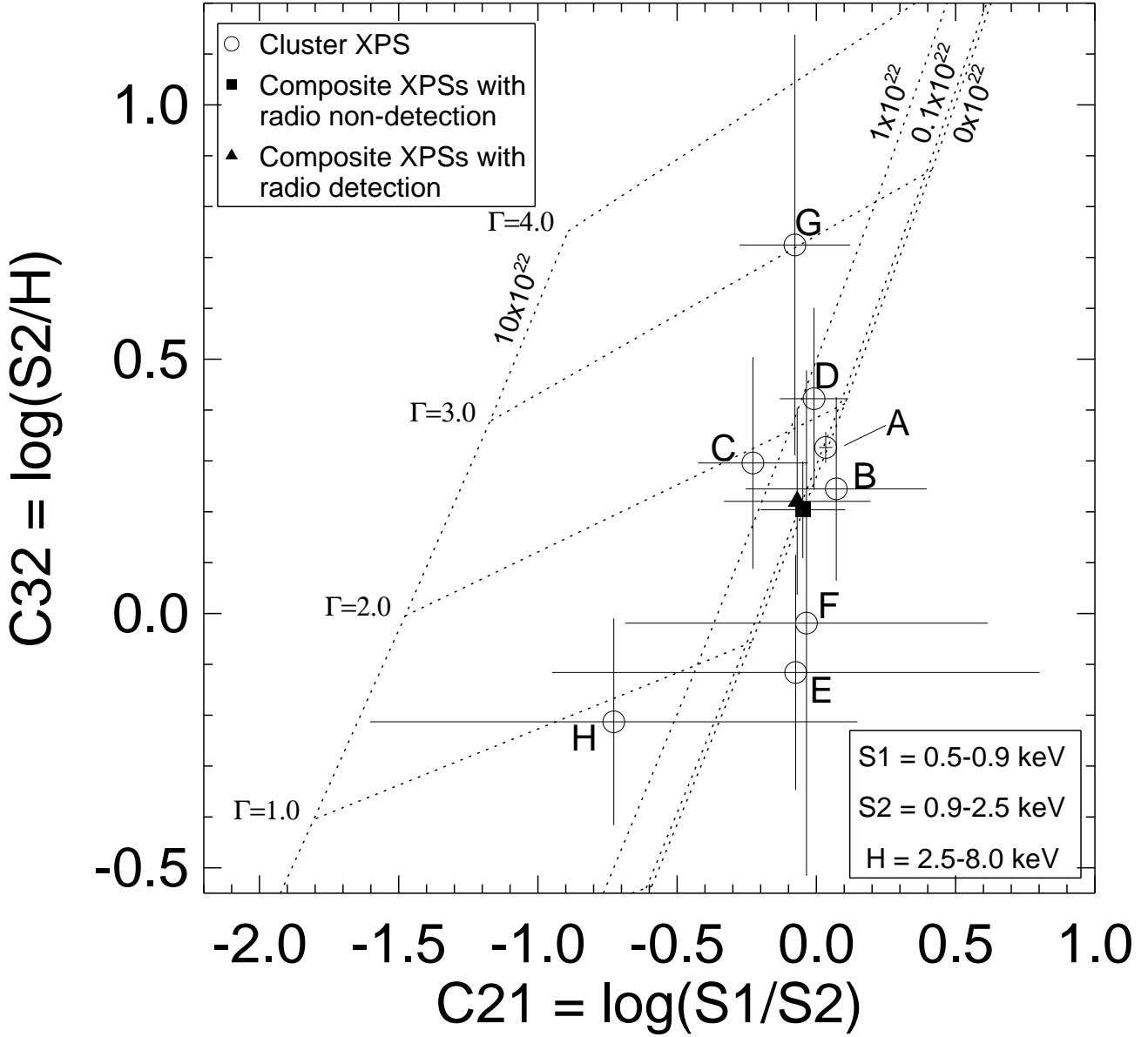}
\caption[scale=0.5]{\scriptsize{
Rest-frame X-ray colors of cluster X-ray point sources.  The colors are defined as follows:
S1=0.5-0.9 keV, S2=0.9-2.5 keV, and H=2.5-8.0 keV.
Overlaid is a grid of the expected color ratios in this color system for
$\Gamma$=1-4 and n$_H$=0,0.1,1.0,10$\times10^{22} cm^{-2}$.  The open circles are our detected cluster XPSs.  
Letter identifiers correspond to the X-ray point sources listed in Table~\ref{tab:xps}.
The filled square is the composite color of the stacked XPS spectra {\it without} radio emission above our limits,
excluding the one XPS (Abell 963 X1, Object A) with a typical Seyfert optical emission line spectrum.  The filled triangle
is the composite color of the stacked XPS spectra {\it with} radio emission above our limits.
The X-ray colors of the stacked XPS spectra are consistent with an unobscured AGN with $\Gamma=1.7$.  
}}
\label{fig:xray_color}
\end{center}
\end{figure}

\begin{figure}
\begin{center}
\includegraphics{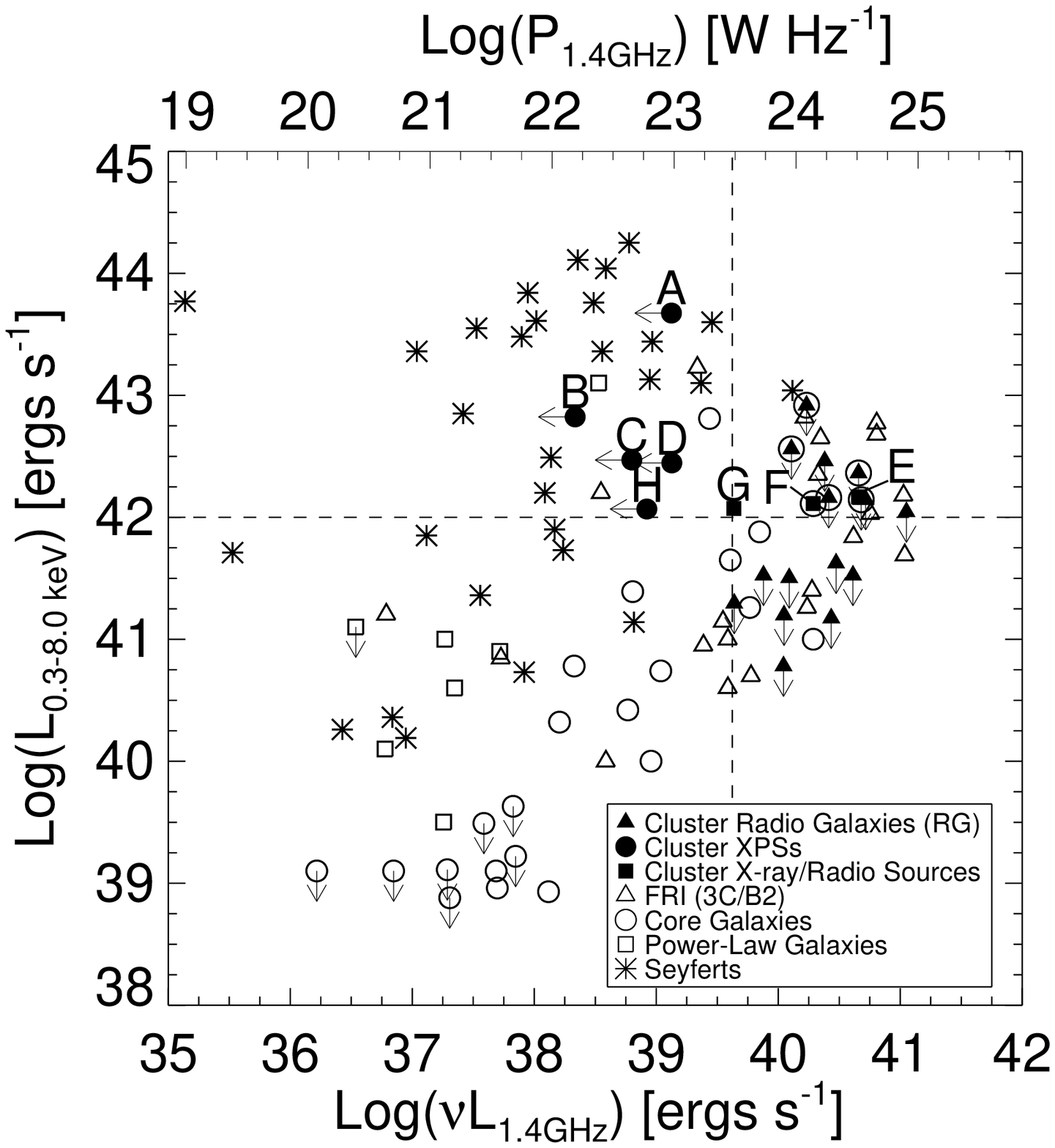}
\caption[scale=0.5]{\scriptsize{
X-ray (0.3-8.0keV) versus radio (1.4 GHz) luminosity for cluster radio galaxies and X-ray point sources.
Cluster sources are compared to typical FR 1s \citep{2004ApJ...617..915D},
ellipticals classified as ''core" and ''power-law" galaxies \citep{2006A&A...447...97B}, and
Seyferts \citep{2007A&A...469...75C}.
The vertical and horizontal dashed lines represent our radio power limit of P$_{1.4GHz}>3\times$10$^{23}$ W Hz$^{-1}$ and
X-ray luminosity limit (0.3-8.0 keV) of L$_{X}>10^{42}$ ergs s$^{-1}$.
Cluster radio galaxies are identified with {\it filled triangles} (radio-loud BCGs are additionally identified with a black circle around the
triangle), while cluster X-ray point sources are identified with {\it filled circles}.
Letter identifiers correspond to X-ray point sources listed in Table~\ref{tab:xps}.
Filled squares are sources with both detectable X-ray and radio emission within our defined luminosity limits (with BCGs
identified as above).
Notice that our X-ray point sources with radio non-detections and our radio galaxies with X-ray non-detections do
not overlap in this plot, suggesting two different populations of objects.
}}
\label{fig:lum_ratio}
\end{center}
\end{figure}

The detailed optical properties of Abell 963 X2 (source D in Figures~\ref{fig:cmd}, \ref{fig:xray_color}, \& \ref{fig:lum_ratio}) 
also support the HBL classification of the XPSs.
Figure~\ref{fig:cmd} shows that this XPS host galaxy is $\sim$~0.5 mag bluer in (g-r) than its associated CRS. 
Further to the blue, the discrepancy
enlarges with Sloan (u-g)=0.64$\pm$0.26 while the CRS galaxies in Abell 963 have Sloan (u-g)=1.4-2.4. 
Deeper (u,g,r) imaging of Abell 963 X2, obtained by us at the APO 3.5m telescope, confirms its anomalous colors 
blueward of the Ca II break.  
The radio/XPS source MS1455 X2/R1 (optical object E in Figs.~\ref{fig:cmd}, \ref{fig:xray_color}, \& \ref{fig:lum_ratio}) 
has colors slightly bluer than the CRS, and so could also possess non-thermal emission in the blue.
Table~\ref{tab:blue_xps_colors} compares the observed colors of these two ``blue" XPSs to the
expected colors for a cluster galaxy of similar r-band luminosity and redshift.

This excess blue flux also is confirmed by our
optical DIS spectrum which shows a slightly diminished CaII H\&K ``break strength'' \citep{1987AJ.....94..899D} of 0.4 similar to other
low-luminosity HBLs \citep{1999ApJ...516..145R}. 
Since this is the lowest luminosity host galaxy among our XPS population,
this is the most likely galaxy for which the blue power-law continuum of a BL Lac would be expected to be visible. 
For the other XPSs in our sample, the CaII H\&K ``break strength'' ranges between 0.4--0.5, in support of the low-luminosity BL Lac hypothesis.
Confirmation of
this hypothesis can be obtained by detecting optical polarization in the B-band (blueward of Ca II H\&K at this redshift in Abell 963 X2).
Thus, the observed properties of the XPSs match those of HBLs the best of any known AGN class.

The X-ray luminosity functions obtained for HBLs by \citet{1991ApJ...380...49M} and \citet{2000AJ....120.1626R} show no signs of flattening
down to L$_x$=10$^{43.5}$ ergs s$^{-1}$ (although theoretical luminosity functions of beamed sources all have a flattening at low power
levels; \citealt{1984ApJ...280..569U}) and so numerous lower power HBLs would be expected in bright cluster ellipticals. 
In Figure~\ref{fig:lum_ratio}, excepting the one Seyfert (Object A),
the other XPSs are 1.5--2 orders of magnitude more luminous in X-rays than the sequence of FR 1s and compact core galaxies. Thus, the approximate 
``excess L$_x$'' of the XPSs in our sample above the FR 1 sequence suggests a Doppler-boosting of X-rays by factors of 30--100. 
Taking the simplified model of HBLs described in \citeauthor{1995PASP..107..803U} \citeyearpar{1995PASP..107..803U}, 
our line-of-sight is located well off the radio
beaming axis (thus no beamed radio emission is observed) but within the $\gamma \geq$ 3 X-ray beam. In this case the maximum kinematic
Doppler factor expected for the X-rays is $\delta \sim$~2$\gamma$ and the ratio of observed-to-intrinsic X-ray luminosity is
$\delta^p$ where p=3+$\alpha$ for discrete emitting ``blobs'' and p=2+$\alpha$ for a continuously emitting structure (where $\alpha$ is
the intrinsic spectral index in energy units for the source; here we conservatively assume $\alpha$= 1). In the simplifying assumptions
of the HBL model we obtain a maximum luminosity boosting of 200-1300 depending upon the physical structure of the emitting region (or
even higher if $\Gamma$ is greater than as estimated from the inferred opening angles of 25--60 degrees based upon source counts;
\citealt{2000AJ....120.1626R}). Since
the most luminous HBLs have L$_x \sim$~10$^{44-45}$ ergs s$^{-1}$, the unbeamed X-ray emission is estimated to be 10$^{40-41}$ ergs
s$^{-1}$, or even lower depending upon how these quantities scale with source luminosity. Thus, our estimate above that the XPSs are
boosted in their X-ray emission by 30--100 times seems plausible. 

% new paragraph has been added
The most luminous HBLs are found in large-solid-angle ``serendipitous'' X-ray surveys and so are
quite rare ($\sim$~10$^{-7}$ sources Mpc$^{-3}$ at L$_x \geq$ 10$^{44}$ ergs s$^{-1}$ from the
\citet{2000AJ....120.1626R} XLF). A power-law extrapolation of that XLF yields $\sim$~10$^{-5}$ BL Lacs
per Mpc$^{-3}$ at L$_x \geq$ 10$^{42}$ ergs s$^{-1}$. Since the elliptical galaxy LF finds 
$\sim8\times 10^{-4}$~Mpc$^{-3}$ for L$\geq$L$^*$ \citep{1994AJ....108..437M,1996ApJ...461L..79I}, the expectation
from observed space densities is $\sim$~1 BL Lac in 80~$\geq$ L$^*$ elliptical
galaxies. In this survey we have observed $\sim$~665 bright cluster
ellipticals and found 7 XPSs (not include Abell 963 X1) which translates to $\sim$~1 XPS per 95 bright ellipticals,
remarkably identical to the expectations given the rather large uncertainties
due to counting statistics in both our sample and the \citet{2000AJ....120.1626R} HBL sample.

In conclusion, these XPSs possess all of the qualities (including space densities) expected for 
low luminosity HBLs and thus are related to the higher power radio galaxies in our sample
in that both classes are radio-loud and possess relativistic jets that can carry energy and material into the surrounding ICM. If 
the identification of these XPSs as HBLs is correct, then both the radio-selected and the X-ray selected
AGN in clusters are radio-loud sources, and there is no need to create a new category of AGN for similar sources seen in other
surveys (e.g., deep XMM and {\it Chandra} serendipitous source surveys; \citealt*{2001AJ....121..662B}).

A final known category of X-ray emitter into which the XPSs might be placed is an accreted ``cool core'' thermal corona in galaxy-sized
halos \citep{2007ApJ...657..197S}.  Thermal X-ray coronae of early type galaxies have been known since the X-ray observations of
{\it Einstein} \citep{1985ApJ...293..102F}. 
Examples of this phenomenon have been seen in nearby clusters \citep[e.g.,][]{2002ApJ...578..833Y, 2005ApJ...633..165S}, 
and not necessarily in the BCG. 
We expected to spatially resolve this type of emitting region since the nearest XPSs in our sample are at a distances where the 1 arcsec 
{\it Chandra} resolution is $\sim$~3 kpc. 
But, the surrounding diffuse hot ICM emission may make these cores more difficult to resolve at higher redshift, 
especially when they are detected at such low count levels, so this test is not definitive for our data. 
Additionally, simulations of cool cores in in-falling cluster ellipticals are 1--2 orders of magnitude less than 
the X-ray emission in these XPSs \citep[e.g.,][]{2001ApJ...555L..87V,2007ApJ...657..197S}. 
And while the X-ray spectra of the XPSs we have found can be fit by thermal models with T$_x$ less than the
surrounding medium, a power-law fit is better in all individual cases
and composites (see \S~\ref{subsec:xray_properties}). The typical observational signature for
soft thermal coronae is the presence of the iron L-shell bump at 0.8-1
keV, resulting in a poor power law fit \citep[e.g.,][]{2007ApJ...657..197S}. The one possibility for which we can
imagine creating cool cores as luminous as $\geq$10$^{42}$ ergs s$^{-1}$ is a remnant core of a group BCG that is in the process of falling 
into the cluster. But none of the XPSs we detect (except one cluster BCG which also has radio emission) is in a very bright CRS galaxy. 
We conclude that it is unlikely that these sources are remnant cool cores and that the best model for XPSs is HBL-type BL Lac Objects.

% ********************************  Discussion ***************************************
% *************************************************
\section{Discussion}
\label{sec:discussion}
% *************************************************

Based upon the observed properties of the 20 radio galaxies and the 7 XPSs (total of 24 AGN counting overlaps and
excluding Abell 963 X1)
in these 11 low-$z$ clusters, we conclude that all these sources are radio-loud AGN. The radio galaxies have
properties consistent with high luminosity FR~1 sources while the XPSs without radio emission are most consistent 
with being low power FR~1s with X-ray emission that has been Doppler-boosted above our detection threshold.
Thus, there is observational evidence that both AGN classes (X-ray and radio selected) possess relativistic jets 
that can heat the surrounding ICM through jet interaction. 

Since both classes have luminosity function determinations from previous work, we can extrapolate downward in both X-ray 
and radio flux below our threshold using previous data. We have normalized these extrapolations using our own sample of $\sim$~265 
L $>$ L$^*$ cluster ellipticals at projected radial distances $<$~500 kpc from the cluster ICM X-ray emission centroid, of which 6\%
(16 objects) are radio galaxies at L$_{1.4GHz} >$3 $\times$ 10$^{23}$ W Hz$^{-1}$ and 1\% (3, excluding A963 X1) are 
XPSs at L$_x >$ 10$^{42}$ ergs s$^{-1}$.

% paragraph has been re-written
Based on Figure 4 of \citet{2000A&A...360..463M}, we integrate the AGN radio luminosity function (RLF) 
for log(P)$>$23.4 (high power) and for 21.4$<$log(P)$<$23.4 (low power) and expect 13 times as many 
low power radio galaxies as high power radio sources in our sample (or $\sim$210 low power radio sources).
Based upon these values, we predict that $\sim$ 85\% of all bright CRS galaxies within 500 kpc of cluster center possess radio sources 
with L$_{1.4GHz} >$ 10$^{21.4}$ W Hz$^{-1}$.
The XLF of HBLs predicts even a higher number, all (100\%) bright cluster ellipticals should possess an XPS at 
L$_x \geq$ 10$^{40}$ ergs s$^{-1}$. But the XLF of \citet{1991ApJ...380...49M} and \citet{2000AJ....120.1626R} extends at constant logarithmic 
slope down only to 10$^{43.5}$ ergs s$^{-1}$. 
Since flattening of this function is expected at or below 10$^{43}$ ergs s$^{-1}$ \citep{1984ApJ...280..569U}, the number of predicted HBLs
would be somewhat less than 100\% and so in good agreement with the extrapolation of the RLF. Thus, the two extrapolations
roughly agree in predicting that almost all bright CRS galaxies ($\geq$ L$^*$) at projected radial distance $<$~500 kpc contain radio-loud 
and X-ray loud AGN at L$_{1.4GHz}>10^{21.4}$W Hz$^{-1}$ and L$_x\geq10^{40}$ \xrayunits, respectively (see Fig.~\ref{fig:lum_ratio}). 
This agreement is additional evidence that 
our identification of the XPSs as low luminosity BL Lacs is correct. 
This result suggests that deeper X-ray surveys of nearby clusters will discover large numbers of XPSs in bright ellipticals 
and very deep {\it Chandra} imaging of the Perseus Cluster appears to show just that
\citep{2007MNRAS.382..895S}.

What impact do all of these AGN have on cluster heating models? 
They certainly can provide a distributed source of heat which moves about
the inner 0.5 Mpc or more of these clusters on a characteristic crossing time of a few $\times$ 10$^8$ yrs. 
But is their input miniscule compared with the AGN in the BCG? To
evaluate the heating by these AGN we have assumed results from two recent studies of the relationship between radio luminosity and jet
power by \citet{1995ApJS..101...29B} and by \citet{2004ApJ...607..800B,2008ApJ...686..859B}. First we assume that the magnetic field 
strength in the inner regions of these clusters is relatively 
constant so that the ICM heating rate is proportional only to the AGN radio luminosity (i.e., a constant efficiency factor
for heating the ICM; \citealt{1995ApJS..101...29B}). This assumption allows us to use the observed RLF to predict that all cluster AGN below 
L$_{1.4GHz} \leq$ 3 $\times$ 10$^{23}$ W Hz$^{-1}$ account for $\leq$10\% of cluster heating; whereas the brightest two or three 
sources (including the BCG) contribute $\geq$90\% of the total heating in proportion to their observed radio fluxes. 

However, a recent re-evaluation of the relationship between jet power and radio luminosity has been undertaken by 
\citet{2004ApJ...607..800B,2008ApJ...686..859B} who find evidence that the 
magnetic field strength is not constant for these sources so that a shallower scaling is required: P$_{jet}$ $\propto$ L$_r^{0.35-0.7}$. 
The largest uncertainty in these calculations appears to be due to source aging; the correlation for the youngest sources is at the steep
end of the range \citep{2008ApJ...686..859B}. Therefore, statistical results obtained 
from the RLF is more robust than a ``snapshot'' of a single cluster's AGN population at any given time.
For example, while our modest-sized sample provides a mean radio galaxy census of
$\sim$~2 per cluster at L$_{1.4GHz} \geq$ 3 $\times$ 10$^{23}$ W Hz$^{-1}$, the spread in relative radio luminosities and thus ICM heating in
these clusters is large. For example, 5 of these clusters have their total radio luminosity dominated by a BCG radio galaxy, 5 do not and
one has no radio galaxy above our flux limit at all. Nevertheless, if the \citet{2008ApJ...686..859B} scaling is correct, weaker radio galaxies 
are proportionally more important than previously believed in heating the ICM. If P$_{jet}$ $\propto$ L$_r^{0.5}$, 
sources weaker than 3 $\times$ 10$^{23}$ W Hz$^{-1}$ account for more than half ($\sim$~55\%) of the heating and so must be taken into account in the modeling. 
This spatially-distributed (and moving!) heat input does however solve the problem of heat distribution throughout the ICM.

From the perspective of AGN heating models, the radio galaxy data in Table~\ref{tab:radio_galaxies} take on a new importance. 
Approximately half of these clusters have luminous radio sources in their BCGs (the
radio galaxy MS1008-R3 is in a galaxy $\sim$~1 mag fainter than the BCG despite being somewhat close to the ICM emission centroid) and
one cluster (Abell 2111) has no luminous radio galaxies at all. If the AGN heating model for clusters is correct, we might expect quite
a wide variation in temperature profiles and/or temperature variations at different radii for clusters with and without luminous BCG
radio sources. It is also possible that peculiar motions of these heat sources within clusters may smooth-out these differences. 
What is clear from the radio luminosities in Table~\ref{tab:radio_galaxies} is that the luminous radio sources in non-BCGs cannot be ignored in cluster 
heating models. 

% **********  Conclusions *************************
% *************************************************
\section{Conclusion}
\label{sec:conclusions}
% *************************************************
We have presented a summary of first results from a large survey of X-ray and radio-selected AGN in clusters of galaxies.
In the 11 clusters we chose as potential progenitors to the present-day Coma cluster, we find
20 radio galaxies and 8 X-ray point sources (XPSs) with 75\% of these AGN centrally concentrated within 500 kpc
projected radius. This central concentration of the AGN is significantly different ($>$ 99.999\% significance level) from the 
radial distribution of the bright ($>$L$^*$) cluster red sequence (CRS) galaxies. This extreme central concentration
strongly suggests that an AGN triggering mechanism similar to the Bondi accretion model of \citet{2006MNRAS.372...21A} is operating 
in these clusters. However, since most of these AGN are moving supersonically with respect to the cluster ICM, Hoyle-Littleton
accretion must be operable, making any gas accretion onto a supermassive Black Hole extremely inefficient. Therefore, other models for
triggering and powering these radio-loud AGN should be considered (e.g., extracting Black Hole spin; \citealt*{1977MNRAS.179..433B}).

Except for 3 X-ray/radio sources, cluster radio galaxies and XPSs generally are not coincident within our 
observed luminosity limits of L$_{1.4GHz}\leq3\times10^{23}$ \radiounits\ and L$_{0.3-8.0keV}\geq10^{42}$ \xrayunits, respectively.
While radio galaxies are certainly associated with AGN activity, we questioned prima facie whether cluster 
XPSs in CRS ellipticals are truly AGN. These X-ray sources show no strong evidence for obscuration and 
are hosted by luminous red galaxies with no evidence for typical Seyfert-like emission signatures. But non-AGN models for cluster XPSs 
(e.g., ``cool core'' emission in luminous galaxy nuclei) seem quite implausible, with these XPSs 
being 1--2 orders of magnitude brighter than expected in 
simulations and as observed in very nearby clusters like Virgo \citep{2007ApJ...657..197S}. 

These XPSs are not spatially resolved in the 
{\it Chandra} images and have X-ray spectra which individually or in composite are better fit as
power-law sources, not thermal sources. We conclude that the most viable model for the XPSs is that they are the beamed X-ray emission 
from low-luminosity BL Lac Objects, similar to the high-energy-peaked BL Lacs (HBLs) studied in detail by 
\citet{1985ApJ...298..619S}, \citet{1995ApJ...444..567P}, and \citet{1999ApJ...516..145R}. 
HBLs are found in luminous, passive ellipticals, have steep X-ray spectra with no evidence for
internal absorption, optical spectra which in many cases are indistinguishable from normal giant ellipticals and are found in rich groups 
and moderately rich clusters at similar redshifts to those observed here (see \citealt{1995PASP..107..803U} for a review of their properties). 
The presence of excess blue light (blueward of the Ca~II break) 
when compared to CRS elliptical galaxies in one XPS (Abell 963 X2) supports the HBL classification of these
sources; a $B$-band polarization detection in this object would solidify this assertion. One other XPSs with a smaller amount of excess
blue light also is present in our sample.

X-ray sources with similar properties have been termed X-ray emitting, passive elliptical galaxies or ``optically-dull AGN''
in other contexts \citep[e.g.,][]{2001AJ....121..662B}, 
and so we suggest based on our current observations that these sources are low luminosity HBLs as well.
More accurate colors and calibrated spectra can be used to test this hypothesis by searching for the excess blue light we have found in
at least one of our sample members.
Therefore, both the XPSs and the radio sources are most readily identified as radio-loud AGN with jets which can transfer heat into the
surrounding ICM. Extrapolations of the radio galaxy population in these clusters using the observed RLF of AGN \citep{2000A&A...360..463M} 
and the XPS population 
using the observed XLF of HBLs \citep{2000AJ....120.1626R} converge to predict that 85-100\% of all bright ellipticals within projected 
radii of 500 kpc from rich cluster centers will be detected by deep radio maps (L$_{1.4GHz}$ $\geq$ 10$^{21.4}$ W Hz$^{-1}$) and 
X-ray imaging (detection limits of L$_{0.3-8.0keV}\geq$ 10$^{40}$ ergs s$^{-1}$). 
There is already a hint that this prediction is correct in that \citet{2007MNRAS.382..895S} have found that all bright
ellipticals in the very core of the Perseus Cluster have XPSs (L$_{0.3-8.0keV}=10^{38-40}$\xrayunits) based on a very deep X-ray image of the core of that cluster.

These results have an impact on models for ICM heating. Based upon the RLF for AGN and the jet power vs. radio luminosity scaling law
of \citet{1995ApJS..101...29B}, we expect that the few brightest radio galaxies
(including the radio source associated with the BCG) contribute $\geq$90\% of
the heating while the remainder at L$_{1.4GHz}$ $\leq$ 3$\times$10$^{23}$ W Hz$^{-1}$ contribute $\leq$10\%. But if the more recent scaling
between jet power and radio luminosity advocated by \citet{2008ApJ...686..859B} is correct, the lower power sources 
(below our X-ray and radio detection limits) contribute $\geq$55\% of the heat input.
Due to their relative motion, the non-BCG AGN heat the ICM more uniformly and would not be expected to possess obvious X-ray ``bubbles'' in the 
ICM associated with their jets due to their peculiar velocities. 
It seems likely that these ideas can be used to test whether AGN are responsible for heating the cluster
ICM as has been proposed \citep{2004ApJ...611..158R}. Putting together two of our results, (1) radio galaxies are more
centrally-concentrated than CRS galaxies as a whole and (2) large numbers of low power radio galaxies are responsible for most of the ICM
heating, suggests that a feedback mechanism by which the density of the ICM triggers the heat sources to offset cooling may be operable in
rich clusters of galaxies. 
We might also expect that those clusters whose total radio luminosities are dominated by radio sources in
the BCG might possess a different temperature profile or detailed temperature/density signatures than clusters dominated by non-BCG 
radio galaxies. The existence
of these signatures and what these signatures might be will be addressed in a future publication \citep{hallman_future}.

\acknowledgments
QNH acknowledges the support from the {\it Chandra} Theory/Archive Grant AR8-9013X.
EJH acknowledges the support from NSF AAPF AST-0702923.
QNH acknowledges useful conversations with Erica Ellingson on the details of the CNOC datasets and providing
us with the PPP photometry catalogs for several CNOC clusters.
We would like to thank Brian Keeney for assistance with our optical observations.
JTS acknowledges useful conversations with M. Begelman, C.S. Reynolds, and T. Jeltema.

Funding for the SDSS and SDSS-II has been provided by the Alfred P. Sloan Foundation, the Participating Institutions, the National Science Foundation, the U.S. Department of Energy, the National Aeronautics and Space Administration, the Japanese Monbukagakusho, the Max Planck Society, and the Higher Education Funding Council for England. The SDSS Web Site is http://www.sdss.org/.

This research has made use of data obtained from the {\it Chandra} Data Archive and software provided by the 
{\it Chandra} X-ray Center (CXC) in the application packages CIAO, ChIPS, and Sherpa.

This work made use of images and/or data products provided by the {\it Chandra} Multi-wavelength Project 
\citep[ChaMP; ][]{2004ApJS..150...43G} supported by NASA. Optical data for ChaMP are
obtained in part through the National Optical Astronomy Observatory (NOAO), operated by the Association of Universities
for Research in Astronomy, Inc. (AURA), under cooperative agreement with the National Science Foundation.

This research has made use of the NASA/IPAC Extragalactic Database (NED) which is operated by the 
Jet Propulsion Laboratory, California Institute of Technology, under contract with the National Aeronautics and Space Administration,
as well as NASA's Astrophysics Data System.

{\it Facilities:} \facility{CXO}, \facility{ARC (3.5m)}, \facility{VLA}, \facility{Sloan}

\clearpage

% ***********************************
% Road to Coma plot
% ***********************************

% ***********************************
% CMD plot
% ***********************************

% ***********************************
% Radial Distribution plot
% ***********************************

% ********** X-ray Color PLOT *************

% ********** X-ray vs Radio Lum. Plot *************

\clearpage

\begin{landscape}

% ----------------------------------------------
% TABLE: General Observations
% ----------------------------------------------
\begin{deluxetable*}{lccccccccccc}
%\begin{deluxetable}{lccccccccccc}
%\rotate
\tabletypesize{\scriptsize}
\tablecaption{``Road to Coma'' Cluster Sample for $z<0.4$ \label{tab:data_obs}}
\tablewidth{0pt}
\setlength{\tabcolsep}{0.03in}
\tablehead{
\colhead{Cluster} & \colhead{{\it z}} & \colhead{{\it Chandra}} & \colhead{ACIS} & \colhead{Exp. Time} & \colhead{T} & \colhead{F$_{X,limit}$} & \colhead{L$_{X,limit}$} & \colhead{VLA} & \colhead{S$_{1.4 GHz,limit}$} & \colhead{P$_{1.4 GHz,limit}$} & \colhead{Comments} \\
\colhead{Name} &  \colhead{} & \colhead{ObsID} & \colhead{chip} & \colhead{(ksec)} & \colhead{(keV)} &\colhead{$\times$10$^{-15}$} & \colhead{$\times$10$^{41}$} & \colhead{1.4 GHz Map} & \colhead{(mJy)} & \colhead{$\times 10^{23}$} & \colhead{} \\
\colhead{(1)} &  \colhead{(2)} & \colhead{(3)} & \colhead{(4)} & \colhead{(5)} & \colhead{(6)} & \colhead{(7)} & \colhead{(8)} &\colhead{(9)} & \colhead{(10)} & \colhead{(11)} &\colhead{(12)}
}
\startdata
MS 0440.5+0204  & 0.197 & 4196  & S3 & 38.2     & 6.1$^{+0.6}_{-0.5}$   & ~2.2 & ~2.4  & AS873  & 0.4   & 0.4   & AI72, A-array \\
Abell 963       & 0.206 & 903   & S3 & 35.8     & 7.1$^{+0.2}_{-0.2}$   & ~1.7 & ~2.1  & FIRST  & 1.1   & 1.4   & \\
RX J0952.8+5153 & 0.214 & 3195  & S3 & 25.6     & 5.2$^{+0.2}_{-0.2}$   & ~2.3 & ~3.1  & FIRST  & 0.4   & 0.5   &  \\
Abell 2111      & 0.229 & 544   & I3 & ~9.7     & 7.2$^{+0.6}_{-0.5}$   & ~8.2 & 12.7$^{*}$  & AS873  & 0.3   & 0.5   & FIRST \\
MS 1455.0+2232  & 0.260 & 4192  & I3 & 75.6     & 4.5$^{+0.1}_{-0.1}$   & ~2.0 & ~4.1  & AS873  & 0.4   & 0.8   & FIRST \\
Abell 1758      & 0.280 & 2213  & S3 & 35.8     & 7.5$^{+0.5}_{-0.3}$   & ~1.9 & ~4.8  & FIRST  & 0.4   & 1.0   & \\
MS1008.1-1224   & 0.306 & 926   & I3 & 33.4     & 6.1$^{+0.4}_{-0.5}$   & ~3.2 & ~9.7  & \citet{1999AJ....117.1967S}    & 0.6           & 1.8    &  \\
MS2137.3-2353   & 0.313 & 5250  & S3 & 23.5     & 4.6$^{+0.1}_{-0.1}$   & ~2.2 & ~7.0  & AB1022 & 0.2   & 0.6   &  \\
Abell 1995      & 0.319 & 906   & S3 & 55.4     & 8.8$^{+0.5}_{-0.5}$   & ~1.2 & ~3.8  & AS873  & 0.5   & 1.6   & FIRST \\
MS1358.4+6245   & 0.330 & 516   & S3 & 34.1     & 8.2$^{+0.7}_{-0.7}$   & ~1.5 & ~5.2  & FIRST	& 0.5$^{**}$   & 1.7$^{**}$ & \citet{1999AJ....117.1967S} \\
Abell 370       & 0.373 & 515   & S3 & 59.3     & 8.5$^{+0.5}_{-0.5}$   & ~1.2 & ~5.8  & FIRST  & 0.4   & 1.9   & AL159, A-array\\
\enddata
\tablecomments{Columns:
(1) cluster name
(2) cluster redshift
(3){\it Chandra} observation ID
(4) ACIS aimpoint
(5) Exposure time of {\it Chandra} observation after filtering for flaring events
(6) ICM temperatures determined by XSPEC modeling of spectra extracted with a 1 Mpc (projected) radius and
with cooling cores excised in some clusters.  See \S~\ref{subsubsec:xray_data_icm} for details.
(7) Flux limit (0.3-8.0 keV) in ergs cm$^{-2}$ s$^{-1}$ for a point source near the edge of our survey region (R $=$ 1 Mpc).  See
\S~\ref{subsubsec:xray_data_xps} for more details.
(8) Luminosity limit (0.3-8.0 keV) in ergs s$^{-1}$ for the flux limits in Column 7
(9) VLA program or Reference; AS873 Observations were obtained in B or C-array configuration
(10-11) 1.4 GHz Radio flux density in mJy and power limits in W Hz$^{-1}$ for a 3$\sigma$ detection at the edge of our survey region
(12) Additional VLA programs used to identify radio sources
}
\tablenotetext{*}{The X-ray luminosity limit in Abell 2111 is slightly greater than our minimum threshold of L$_{X}>1\times10^{42}$ \xrayunits.  See \S~\ref{subsubsec:xray_data_xps} for our justification on its inclusion in the cluster sample.}
\tablenotetext{**}{These limits are extrapolated from 5 GHz observations assuming F$_{\nu}\propto\nu^{-\alpha}$ and $\alpha$=0.7. See \S~\ref{subsec:radio_data} for details.}
%\end{deluxetable}
\end{deluxetable*}

\clearpage

% ----------------------------------------------
% TABLE: Cluster XPS Properties
% ----------------------------------------------
\begin{deluxetable*}{lcclcccccl}
%\begin{deluxetable}{lcclcccccl}
%\rotate
\tabletypesize{\scriptsize}
\tablecaption{Cluster X-ray Point Sources with L$_{0.3-8.0 keV} > 10^{42}$ ergs s$^{-1}$ \label{tab:xps}}
\tablewidth{0pt}
\setlength{\tabcolsep}{0.03in}
\tablehead{
\colhead{Object} & \colhead{$\alpha$} & \colhead{$\delta$} & \colhead{{\it z}} & \colhead{r} & \colhead{Radius} & \colhead{Net Counts} & \colhead{F$_{X,observed}\times10^{-15}$} & \colhead{L$_{X,rest}\times10^{42}$} &
\colhead{Comments}\\
\colhead{} & \colhead{(J2000)} & \colhead{(J2000)} & \colhead{}  & \colhead{mag} & \colhead{(h$^{-1}_{70}$ Mpc)} & \colhead{(0.3-8.0 keV)} & \colhead{(ergs cm$^{-2}$ s$^{-1}$)} & \colhead{(ergs s$^{-1}$)} & \colhead{} \\
\colhead{(1)} &  \colhead{(2)} & \colhead{(3)} & \colhead{(4)} & \colhead{(5)} & \colhead{(6)} & \colhead{(7)} & \colhead{(8)} &\colhead{(9)} & \colhead{(10)}
}
\startdata
(A) A963-X1     & 10:17:00.8    & +39:04:33.0   & 0.209$^a$     & 18.22 & 0.360 & 2150.0 $\pm$ 48.8     & 394.2 $\pm$ 8.9  & 45.8 $\pm$ 1.0 & [O III], H$\alpha$ emission lines\\
(B) MS1008-X3   & 10:10:35.3    & -12:40:22.1   & 0.3095$^b$    & 19.39 & 0.264 & ~~60.2 $\pm$ ~9.5     & ~23.4 $\pm$ 3.7  & ~6.5 $\pm$ 1.0 & \\
(C) A370-X6     & 02:40:00.2    & -01:32:34.1   & 0.380$^{ac}$  & 19.94 & 0.831 & ~~62.3 $\pm$ ~9.5     & ~~6.5 $\pm$ 1.0  & ~2.9 $\pm$ 0.4 & \\
(D) A963-X2     & 10:16:58.7    & +39:02:12.7   & 0.206$^a$     & 20.74 & 0.237 & ~115.1 $\pm$ 12.8     & ~23.8 $\pm$ 2.7  & ~2.8 $\pm$ 0.3 & Passive spectrum with blue excess\\
(E) MS1455-X2   & 14:56:58.7    & +22:18:46.3   & 0.2582$^b$    & 20.05 & 1.008 & ~~26.0 $\pm$ ~6.5     & ~~7.7 $\pm$ 1.9  & ~1.5 $\pm$ 0.4 & Coincident with radio source, MS1455-R1; \\
                &               &               &               &       &       &                       &               &       & Probable blue excess \\
%(F) A1758-X13   & 13:32:59.9    & +50:33:44.5   & 0.279     & 19.82 & 0.637 & ~~31.1 $\pm$ ~7.5     & ~~5.8 $\pm$ 1.4  & ~1.3 $\pm$ 0.3 &  Probable blue excess \\
(F) A1758-BCG   & 13:32:38.3    & +50:33:36.2   & 0.279$^a$     & 17.19 & 0.000 & ~~30.9 $\pm$ ~9.9     & ~~5.7 $\pm$ 1.8  & ~1.3 $\pm$ 0.4 & Coincident with radio source, A1758-R4 \\
(G) MS0440-X8   & 04:43:14.2    & +02:12:04.7   & 0.1996$^b$    & 18.37$^{*}$   & 0.403 & ~~48.6 $\pm$ ~8.4     & ~11.0 $\pm$ 1.9  & ~1.2 $\pm$ 0.2 & Coincident with radio source, MS0440-R5 \\
(H) RXJ0952-X6  & 09:52:36.2    & +51:56:32.9   & 0.22$^a$      & 19.47 & 0.842 & ~~33.4 $\pm$ ~7.8     & ~~8.7 $\pm$ 1.9  & ~1.1 $\pm$ 0.2 & \\
\enddata
\tablecomments{Columns: (1) Cluster X-ray point source IDs, with same letter ID as displayed in Figs.~\ref{fig:cmd}, \ref{fig:xray_color} \& \ref{fig:lum_ratio}.
(2-3) (J2000) RA \& DEC of X-ray point source
(4) redshift (5) observed Sloan r-band magnitude, except for MS0440-X8 which is Gunn r-band magnitude (6) projected radial distance (Mpc) from the peak in the cluster X-ray emission
(7) net counts in 0.3-8 keV bandpass (8) 0.3-8.0 keV Flux (9) 0.3-8.0 keV Rest-frame Luminosity assuming a power-law spectrum with a photon index of 1.7 (N$_E\propto$ E$^{-\Gamma}$).}
\tablenotetext{*}{Gunn r-magnitude}
\tablenotetext{a}{This work}
\tablenotetext{b}{CNOC; \citet{1998ApJS..116..211Y}}
\tablenotetext{c}{\citet{2001AJ....122.2177B}}
%\end{deluxetable}
\end{deluxetable*}

\clearpage

 ----------------------------------------------
% TABLE: Cluster Radio Galaxy Properties
% ----------------------------------------------
\begin{deluxetable*}{lccllccclccr}
%\begin{deluxetable}{lccllccclccr}
%\rotate
\tabletypesize{\scriptsize}
\tablecaption{Cluster Radio Galaxies with P$_{1.4GHz}>3\times10^{23}$W Hz$^{-1}$ \label{tab:radio_galaxies}}
\tablewidth{0pt}
\setlength{\tabcolsep}{0.05in}
\tablehead{
\colhead{} & \colhead{$\alpha$} & \colhead{$\delta$}  & \colhead{{\it z}} & \colhead{r} & \colhead{(g-r)} & \colhead{(r-i)} & \colhead{CRS} & \colhead{Radius} & \colhead{S$_{1.4GHz}$} & \colhead{P$_{1.4GHz}\times10^{23}$} & \colhead{L$_X\times$10$^{42}$} \\
\colhead{Object} & \colhead{(J2000)} & \colhead{(J2000)} & \colhead{}  & \colhead{mag} & \colhead{color} & \colhead{color} & \colhead{color} & \colhead{Mpc} & \colhead{(mJy)} & \colhead{(W Hz$^{-1}$)} & \colhead{ergs s$^{-1}$} \\
\colhead{(1)} & \colhead{(2)} & \colhead{(3)}   &\colhead{(4)} & \colhead{(5)} &\colhead{(6)} & \colhead{(7)} &\colhead{(8)} &\colhead{(9)} & \colhead{(10)} & \colhead{(11)} & \colhead{(12)}
}
\startdata
MS0440-R3*      & 04:43:10.0    & +02:10:18.4   & 0.199$^a$     & 17.03$^{**}$  & 0.99 && 1.00 $\pm$ 0.30  & 0.001 & 30.2~ $\pm$ 0.07 & ~34.2 $\pm$ 0.1   & $<$ 1.4 \\
MS0440-R4       & 04:43:09.4    & +02:10:00.0   & 0.193$^a$     & 18.67$^{**}$          & 0.93& & 1.00 $\pm$ 0.30  & 0.062 & ~7.5~ $\pm$ 0.08 & ~~8.0 $\pm$ 0.1   & $<$ 0.2 \\
MS0440-R5       & 04:43:14.2    & +02:12:04.2   & 0.120$^a$     & 18.37$^{**}$          & 0.92&& 1.00 $\pm$ 0.30  & 0.392 & ~2.7~ $\pm$ 0.2~ & ~~3.1 $\pm$ 0.2    & 1.2 $\pm$ 0.2 \\
A963-R2         & 10:17:22.0    & +39:04:39.1   & 0.197$^c$     & 17.14 & 1.34&& 1.31 $\pm$ 0.21  & 0.806 & ~3.82 $\pm$ 0.38 & ~~4.2~ $\pm$ 0.4   & $<$ 0.2 \\
RXJ0952-R1*     & 09:52:49.1    & +51:53:05.0   & 0.215$^d$     & 16.56 & 1.28&& 1.31 $\pm$ 0.18  & 0.009 & 13.70 $\pm$ 0.13 & ~18.5 $\pm$ 0.2    & $<$ 1.5 \\
RXJ0952-R3      & 09:52:46.9    & +51:54:42.2   & 0.22~$^c$     & 19.48 & 1.44& & 1.31 $\pm$ 0.18  & 0.351 & ~5.54 $\pm$ 0.13 & ~~7.9 $\pm$ 0.2   & $<$ 0.1 \\
MS1455-R1       & 14:56:58.7    & +22:18:46.3   & 0.258$^j$     & 20.05 & 1.32& & 1.47 $\pm$ 0.30  & 1.014 & 19.5 $\pm$ 0.1 & ~39.9 $\pm$ 0.2     & 1.5 $\pm$ 0.4 \\
MS1455-R2*      & 14:57:15.1    & +22:20:34.3   & 0.258$^j$     & 16.50 & 1.24& & 1.47 $\pm$ 0.30  & 0.007 & 15.9 $\pm$ 0.1 & ~32.5 $\pm$ 0.2     & $<$ 2.3 \\
MS1455-R3       & 14:57:08.5    & +22:20:14.5   & 0.25~$^c$     & 18.39 & 1.56& & 1.47 $\pm$ 0.30  & 0.379 & 14.2 $\pm$ 0.1 & ~27.0 $\pm$ 0.2     & $<$ 0.2  \\
A1758-R4*       & 13:32:38.4    & +50:33:35.9   & 0.278$^c$     & 17.19 & 1.55& & 1.52 $\pm$ 0.39  & 0.000 & ~6.03 $\pm$ 0.15 & ~14.6 $\pm$ 0.4   & 1.3 $\pm$ 0.4 \\
A1758-R6        & 13:33:02.0    & +50:29:28.6   & 0.279$^d$     & 17.96 & 1.60& & 1.52 $\pm$ 0.39  & 0.670 & ~1.76 $\pm$ 0.15 & ~~4.3 $\pm$ 0.4   & $<$ 0.9 \\
MS1008-R1       & 10:10:31.1    & -12:37:02.6   & 0.311$^{ce}$  & 18.26$^{**}$          & 1.19& & 1.25 $\pm$ 0.40  & 0.730 & 25.6~ $\pm$ 0.2~ & ~80.3 $\pm$ 0.6   & $<$ 0.3 \\
MS1008-R2       & 10:10:29.4    & -12:38:27.4   & 0.300$^c$     & 18.72$^{**}$          & 1.23& & 1.25 $\pm$ 0.40  & 0.393 & 12.9~ $\pm$ 0.2~ & ~37.2 $\pm$ 0.6   & $<$ 1.1 \\
MS1008-R3       & 10:10:32.3    & -12:39:34.1   & 0.301$^f$     & 18.75$^{**}$          & 1.15& & 1.25 $\pm$ 0.40  & 0.042 & ~5.9~ $\pm$ 0.2~ & ~17.2 $\pm$ 0.6   & $<$ 3.1 \\
MS2137-R4*      & 21:40:15.1    & -23:39:40.0   & 0.314$^{c}$   & 17.36 && 0.57   & 0.50 $\pm$ 0.25  & 0.001 & ~3.8~ $\pm$ 0.5  & ~12.2 $\pm$ 1.6 & $<$ 8.3 \\
A1995-R1        & 14:53:00.5    & +58:03:20.0   & 0.325$^g$     & 19.15 && 0.58   & 0.59 $\pm$ 0.21  & 0.151 & ~8.43 $\pm$ 0.15 & ~29.3 $\pm$ 0.5 & $<$ 0.3 \\
MS1358-R1*      & 13:59:50.6    & +62:31:04.2   & 0.326$^b$     & 17.75 && 0.66   & 0.57 $\pm$ 0.33  & 0.005 & ~2.1~ $\pm$ 0.9~ & ~~7.4 $\pm$ 3.2 & $<$ 3.6  \\
A370-R1         & 02:39:55.3    & -01:34:05.4   & 0.382$^{ch}$  & 19.97 && 0.68   & 0.75 $\pm$ 0.33  & 0.224 & ~4.6~ $\pm$ 0.15 & ~23.3 $\pm$ 0.8 & $<$ 0.4 \\
A370-R2         & 02:39:52.9    & -01:35:01.4   & 0.376$^c$     & 21.21 && 0.88   & 0.75 $\pm$ 0.33  & 0.123 & ~1.51 $\pm$ 0.15 & ~~7.4 $\pm$ 0.7 & $<$ 0.3 \\
A370-R3         & 02:39:56.3    & -01:34:29.3   & 0.370$^h$     & 19.68 && 0.65   & 0.75 $\pm$ 0.33  & 0.280 & ~1.36 $\pm$ 0.05 & ~~6.4 $\pm$ 0.2 & $<$ 0.3 \\
\enddata
\tablecomments{Columns:
(1) Radio Galaxy ID with brightest cluster galaxies (BCGs) denoted by an asterisk.  For Abell 2111, no radio galaxy was detected above our radio power limit.
(2-3) J2000 Coordinates
(4) object redshifts (see reference below)
(5) apparent Sloan r-band magnitude except for those with double asterisks which are Gunn r-band magnitudes
(6) Observed color (g-r) or (7) (r-i)
(8) Mean color of the cluster red sequence. The mean Sloan color and 3$\sigma$ spread of the cluster red sequence for $M^{*}_r<-20.8$.
(9) Projected radial distance from the X-ray peak emission.  Note that the X-ray image of Abell 1758 is a double-peaked in
appearance, indicative of a merging system.  Therefore, the projected radii values are relative to the NW and SE X-ray emission clumps.
(10) observed flux density of the source
(11) observed radio power of the source
(12) Rest frame X-ray Luminosity (0.3-8 keV) assuming a power-law spectrum with $\Gamma$=1.7 (N$_{E}\propto E^{-\Gamma}$).
The BCG X-ray flux limits are calculated assuming a
minimum of 3$\sigma$ above the diffuse cluster emission.  For the remaining radio sources, X-ray flux limits were determined
using net counts (observed minus local background) plus 3$\sigma$ above the background noise.  References for source redshifts:
(a)\citet{1998ApJ...497..573G}
(b) CNOC; \citet{1998ApJS..116..211Y}
(c) This work
(d) SDSS DR6
(e) \citet{1996ApJ...470..172S}
(f) \citet{2007ApJ...664..761M}
(g) \citet{2000ApJ...541...37P}
(h) \citet{1988A&A...199...13M}
(i) \citet{1999AJ....117.1967S}
(j) CNOC; communication from E. Ellingson
}
\tablenotetext{*}{Radio source is hosted in the Brightest Cluster Galaxy}
\tablenotetext{**}{Gunn r-band magnitude}
%\end{deluxetable}
\end{deluxetable*}

\clearpage
\end{landscape}

%\input{agn_fractions}
% ----------------------------------------------
% TABLE: AGN FRACTIONS
% ----------------------------------------------
\begin{deluxetable}{llllllll}
\tabletypesize{\scriptsize}
\tablecaption{AGN Fraction for Cluster Red Sequence (CRS) Galaxies \label{tab:agn_frac}}
\tablewidth{0pt}
\tablehead{
\colhead{Cluster} & \colhead{$N_{CRS}$} & \colhead{$N_{r}$} & \colhead{$f_{R}$} & \colhead{$N_{X}$} & \colhead{$f_{X}$} & \colhead{Color Used} & \colhead{Survey}\\
\colhead{(1)} & \colhead{(2)} & \colhead{(3)} & \colhead{(4)} & \colhead{(5)} & \colhead{(6)} & \colhead{(7)} & \colhead{(8)}
}
\startdata
MS 0440.5+0204  & ~18   & 3$^a$ & 0.167         & 1 & 0.056     & Gunn (g-r)    & CNOC \\
Abell 963       & ~48   & 1     & 0.021         & 2 & 0.042     & Sloan (g-r)   & SDSS DR6 \\
RX J0952.8+5153 & ~23   & 2$^a$ & 0.087         & 1 & 0.044     & Sloan (g-r)   & SDSS DR6 \\
Abell 2111      & ~72   & 0     & 0.000         & 0 & 0.000     & Sloan (g-r)   & SDSS DR6  \\
MS 1455.0+2232  & ~46   & 3$^a$ & 0.065         & 1 & 0.022     & Gunn (g-r)    & CNOC \\
Abell 1758      & ~84   & 2$^a$ & 0.024         & 1 & 0.012     & Sloan (g-r)   & SDSS DR6 \\
MS1008.1-1224   & ~69   & 3     & 0.044         & 1 & 0.015     & Gunn (g-r)    & CNOC  \\
MS2137.3-2353   & ~56   & 1$^a$ & 0.018         & 0 & 0.000     & Sloan (r-i)   & ChaMP \\
Abell 1995      & ~71   & 1     & 0.014         & 0 & 0.000     & Sloan (r-i)   & SDSS DR6  \\
MS1358.4+6245   & ~67   & 1$^a$ & 0.015         & 0 & 0.000     & Sloan (r-i)   & SDSS DR6 \\
Abell 370       & 111   & 3     & 0.027         & 1 & 0.009     & Sloan (r-i)   & APO 2007 \\
                &       &       &               &               &               &       \\
Average         &\ldots &\ldots & 0.044         &\ldots & 0.018 & \ldots        & \ldots \\
Sum             & 665   & 20    & 0.030         & 8 & 0.012     & \ldots        & \ldots \\
\enddata
\tablecomments{Columns: (1) Cluster name (2) number of CRS galaxies with $M^{*}_r<-20.8$ and projected radius $<$ 1 Mpc from cluster center
as estimated using the procedure described in \S~\ref{subsubsec:crs_calc}.
(3) number of radio galaxies with $M^{*}_r<-20.8$ and P$_{1.4GHz}>$ \radiolimit
(4) radio galaxy fraction within our limits
(5) number of X-ray point sources with $M^{*}_r<-20.8$ and L$_{0.3-8 keV}>$ \xraylimit (6) X-ray point source fraction
within our limits (7) observed color used for the CRS galaxy estimation (8) source of the imaging data}
\tablenotetext{a}{The brightest cluster galaxy (BCG) is a radio source.}
\label{tab:agn_fraction}
\end{deluxetable}

%\input{blue_xps_colors}
% ----------------------------------------------
% TABLE: XPS with blue excess
% ----------------------------------------------
\begin{deluxetable}{lccccl}
\tabletypesize{\scriptsize}
\tablecaption{Comparison of Observed vs. Expected Optical Colors for 2 ``Blue" XPSs\label{tab:blue_xps_colors}}
\tablewidth{0pt}
\tablehead{
\colhead{Object} & \colhead{Alternative ID}     & \colhead{Color} & \colhead{Observed} & \colhead{CRS$^{a}$} & \colhead{Comments} \\
}
\startdata
Abell 963 X2    &       Object D        & (u-g)         & 0.70 $\pm$ 0.13       & 1.9 $\pm$ 0.5                 & APO 3.5m      \\
                &                       & (g-r)         & 0.77 $\pm$ 0.06       & 1.26 $\pm$ 0.07               & APO 3.5m      \\
                &                       &               &                       &                               &               \\
MS1455 X2       &       Object E        & (u-g)         & $>$ 0.6               & 2.2 $\pm$ 1.1                 & SDSS DR6      \\
                &                       & (g-r)         & 1.32 $\pm$ 0.08       & 1.45 $\pm$ 0.04               & SDSS DR6      \\
\enddata
\tablenotetext{a}{The expected mean color (SDSS DR6) of a cluster red sequence (CRS) galaxy of
similar luminosity to the object and its associated $1\sigma$ spread about the mean.}
\end{deluxetable}

\end{document}